\documentclass[prl,aps,amsfonts,superscriptaddress,longbibliography,notitlepage,twocolumn]{revtex4-1} 

\usepackage{graphicx}
\usepackage{xcolor}
\usepackage{rotating}
\usepackage{amsmath,amssymb,graphics,amsthm,isomath}
\usepackage{bm}
\usepackage{physics}

\usepackage[colorlinks=true, urlcolor=violet, linkcolor=blue, citecolor=red, hyperindex=true, linktocpage=true]{hyperref}
\usepackage[capitalise,compress]{cleveref}

\makeatletter
\renewcommand{\p@subsection}{}
\renewcommand{\p@subsubsection}{}
\makeatother

\usepackage{xcolor}
\usepackage{mathtools}

\newcommand{\TT}{\mathcal{T}}
\newcommand{\TTT}{\widetilde{\mathcal{T}}}
\newcommand{\RR}{R}
\newcommand{\OO}{\mathcal{O}}
\newcommand{\ii}{\mathrm{i}}
\newcommand{\ee}{\mathrm{e}}
\newcommand{\oo}{\mathrm{o}}
\newcommand{\ddd}[2]{\frac{\mathrm{d} #1}{\mathrm{d} #2}}

\newcommand{\lr}[1]{\left( #1\right)}
\newcommand{\mlr}[1]{\left[ #1\right]}
\newcommand{\mtr}[1]{\mathrm{tr}\left[ #1\right]}
\newcommand{\glr}[1]{\left\{ #1\right\}}
\newcommand{\alr}[1]{\left\langle #1\right\rangle}
\newcommand{\ilr}[1]{\left| #1\right|}

\newcommand{\red}[1]{#1}

\newcommand{\comment}[1]{}
\usepackage{dsfont}

\begin{document}
\title{Universal Entanglement and Correlation Measure in Two-Dimensional Conformal Field Theories}

\author{Chao Yin}
\email{chao.yin@colorado.edu}
\affiliation{Department of Physics and Center for Theory of Quantum Matter, University of Colorado, Boulder, CO 80309, USA}

\author{Zhenhuan Liu}
\email{liu-zh20@mails.tsinghua.edu.cn}
\affiliation{Center for Quantum Information, Institute for Interdisciplinary Information Sciences, Tsinghua University, Beijing 100084, China}

\begin{abstract}
We calculate the amount of entanglement shared by two intervals in the ground state of a (1+1)-dimensional conformal field theory (CFT), quantified by an entanglement measure $\mathcal{E}$ based on the computable cross norm (CCNR) criterion. Unlike negativity or mutual information, we show that $\mathcal{E}$ has a universal expression even for two disjoint intervals, which depends only on the geometry, the central charge $c$, and the thermal partition function of the CFT. We prove this universal expression in the replica approach, where the Riemann surface for calculating $\mathcal{E}$ at each order $n$ is always a torus topologically. By analytic continuation, result of $n=\frac{1}{2}$ gives the value of $\mathcal{E}$. Furthermore, the results of other values of $n$ also yield meaningful conclusions: The $n=1$ result gives a general formula for the two-interval purity, which enables us to calculate the R\' enyi-$2$ $N$-partite information for $N\le 4$ intervals; while the $n=\infty$ result bounds the correlation function of the two intervals. We verify our findings numerically in the spin-1/2 XXZ chain, whose ground state is described by the Luttinger liquid.
\end{abstract}

\date{\today}

\maketitle

\emph{Introduction.}--- It is crucial to understand the structure of entanglement in quantum many-body systems \cite{entan_many}. For critical ground states (and the corresponding low energy sector) described by conformal field theory (CFT), people have derived rigorous results on numerous aspects of entanglement \cite{cft_04,cft_09rev,cft_2d_06,cft_sublead_10,cft_excitation_11,cft_neg_12,cft_excited_12,cft_T_14,cft_corner_15,cft_sym_18}, especially in one spatial dimension with an infinite number of local conformal transformations. Most notably, a single interval of length $\ell$ has a universal entanglement entropy (EE) $S=\frac{c}{3}\ln \ell$ proportional to the central charge $c$ of the CFT \cite{cft_04,cft_09rev}. 

However, for two disjoint intervals $A$ and $B$, it becomes challenging to calculate either EE of them as a whole or the classical correlation and quantum entanglement shared between $A$ and $B$. These quantities would no longer be universal while depending on the full operator content of the CFT \cite{2int_08,2int_info_09,2int_09,2int_11,cft_neg_12}, as we briefly overview below.
Since $A$ and $B$ share a mixed state, there is no unique measure that quantifies the entanglement and correlation in between \cite{entan_rmp}. Two measures have been mainly studied, namely, mutual information \cite{2int_info_09} and positive partial transpose (PPT) negativity \cite{cft_neg_12}. These quantities are calculated in the replica approach, where the R\' enyi version of order $n$ is expressed as a path integral on a Riemann surface composed of $n$ replicas of the system. Unlike the single-interval case, the genus of the Riemann surface grows with $n$ \cite{2int_09,cft_neg_12}. Since CFT calculations on high-genus surfaces become non-universal, the result for general $n$ is very complicated even for free theories \cite{2int_09}, which makes it difficult to analytically continue to the one-replica limit. 
Despite the progress that has been made \cite{2int_Ising_10,2int_XY_10,2int_boson_11,2int_fusion_12,nint_14,2int_num_15,2int_neg_16,2int_spin_16,2int_recursion_18,2int_CoulombGas_21,nint_22,2int_sym_22}, there was no closed-form formula for either entanglement or correlation of two disjoint intervals in general (1+1)-d CFT ground states.

In this Letter, we solve this problem by studying the \emph{computable cross norm (CCNR) negativity} as a different measure for entanglement and correlation. The advantage of this quantity is that the Riemann surface for any number of replicas always has genus $1$, which enables us to draw a connection with \emph{CFT on the torus}, a much better-understood scenario than high-genus surfaces \cite{CFT_book}. By exploiting this quantity at each order $n$ and assuming the thermal free energy of the CFT is known, we derive universal formulas for not only the CCNR negativity, but also other quantifiers of entanglement and correlation in the ground state. These quantifiers include the $2$-interval purity (which generalizes the analytical result in \cite{2int_info_09} to all CFTs), $N$-partite information for up to $N=4$ intervals, and a bound on the correlation function for two intervals. We verify our main results numerically in a spin-$1/2$ XXZ model.

For any state $\rho$ shared by two parties $A$ and $B$, define a realignment matrix $\RR$ with matrix elements 
\begin{equation}\label{eq:RR}
    \bra{a}\bra{a'}\RR\ket{b}\ket{b'}= \bra{a}\bra{b}\rho\ket{a'}\ket{b'},
\end{equation}
where $\glr{\ket{a}}$ and $\glr{\ket{b}}$ are basis for $A$ and $B$, respectively. By definition, $\RR$ is not necessarily a square matrix. It can be proved that $\norm{\RR}_1=\tr(\sqrt{\RR\RR^\dagger})\le 1$ if $\rho$ is separable, so a state is guaranteed to be entangled if $\norm{\RR}_1>1$, so called the CCNR criterion \cite{chen2002matrix}. As a commonly-used mixed-state criterion, it has a similar detection capability as the PPT criterion \cite{collins2016random}. Following the definition of PPT negativity that originates from the PPT criterion \cite{peres1996separability}, the CCNR negativity is defined by
\begin{equation}
    \mathcal{E} = \ln \norm{\RR}_1,
\end{equation}
as an entanglement measure. Note the similarity with the operator entanglement \cite{oper_entan_22,oper_entan_17}.

Now we are ready to set up the problem and present our main result for $\mathcal{E}$. For an infinite 1d system at its ground state described by a 2d CFT, we study the CCNR negativity $\mathcal{E}$ between two intervals $A=[u_a, v_a]$ and $B=[u_b, v_b]$, with $u_a<v_a\le u_b<v_b$ and lengths $\ell_\alpha=v_\alpha-u_\alpha$ ($\alpha=a,b$). We show that $\mathcal{E}$ is related to the torus partition function $Z(\tau/2)$ of the CFT with a \emph{universal} function: \begin{equation}\label{eq:main}
    \mathrm{e}^\mathcal{E} =\frac{ Z\lr{\tau/2}}{\lr{\ell_a\ell_b|u_a-u_b||v_a-v_b||u_a-v_b||u_b-v_a|}^{c/24}},
\end{equation}
where the pure imaginary modular parameter $\tau/2$ of the torus is related to the four-point ratio \begin{equation}\label{eq:x}
    x = \frac{(u_a-v_a)(u_b-v_b)}{(u_a-u_b)(v_a-v_b)} \in (0,1),
\end{equation}
by \begin{equation}\label{eq:xt}
    x= \lr{\frac{\theta_2(\tau)}{\theta_3(\tau)}}^4,
\end{equation}
with $\theta_\nu$ being Jacobi theta functions. \red{$Z(\tau/2)$ is the partition function of the theory on a unit circle at finite temperature $2/|\tau|$, which depends on the CFT model (see Eq.~(S2) in Supplemental Material (SM) \footnote{See Supplemental Material for some derivation details and additional discussions.} for the explicit expression). Other than this dependence, the entanglement measure $\mathcal{E}$ is completely determined by the central charge and geometry. This \emph{universality} also holds for the correlation measures $Z_n$s that we will introduce. We note that \cref{eq:main} is also universal in the strong sense: different microscopic models described by the same CFT will have the same $\mathcal{E}$ in \cref{eq:main} (up to an additive constant). }

\begin{figure}[t]
\centering
\includegraphics[width=0.48\textwidth]{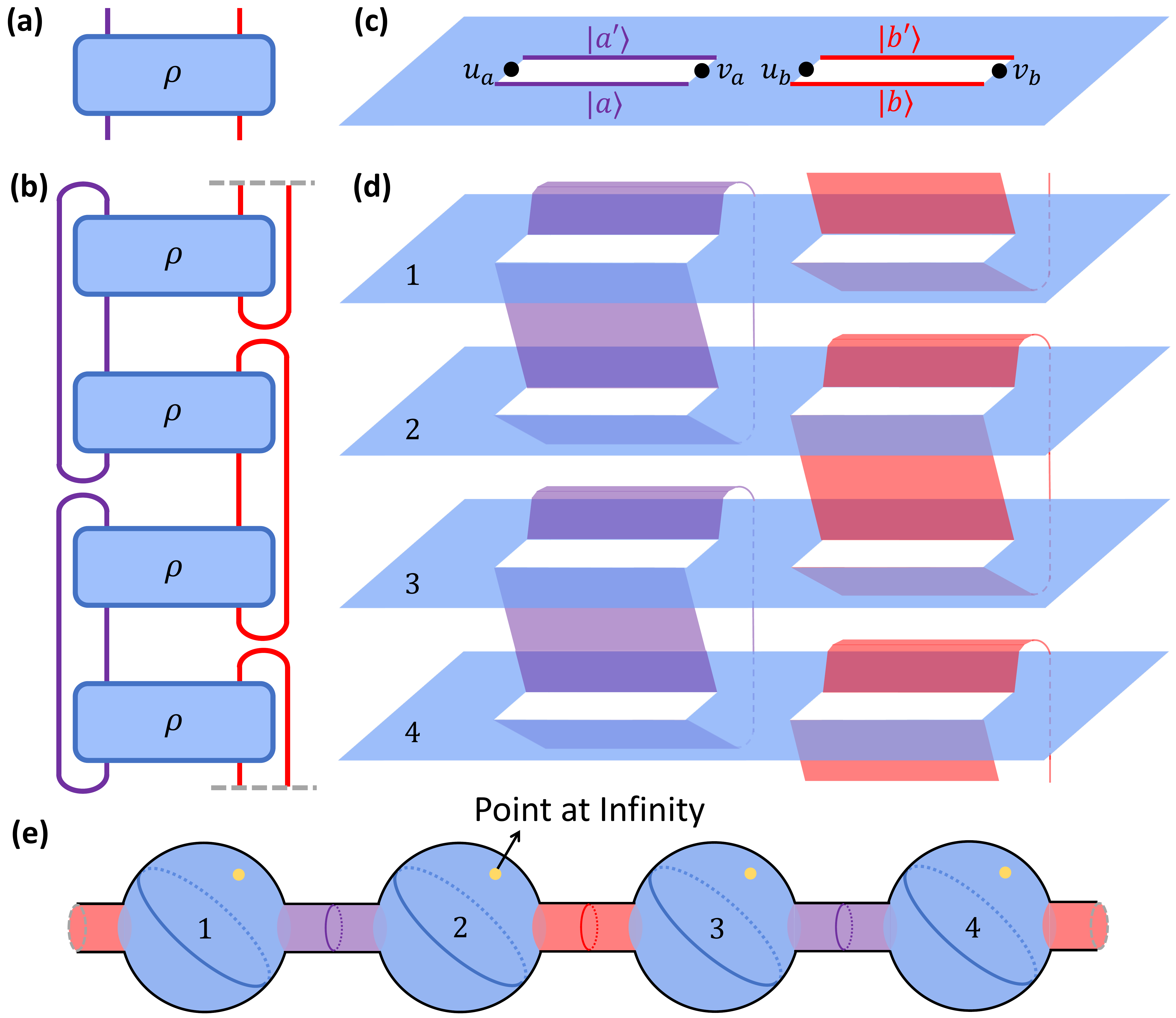}
\caption{\label{fig:Rn} (a) Density matrix $\rho$ as a tensor. (b) Tensor representation of $Z_n$ with $n=2$, where the grey dashed lines represent the periodic boundary condition. (c) Path integral formulation for a matrix element of $\rho$, which is also a matrix element of realignment matrix $R$. (d) The Riemann surface $\mathcal{R}_2$ on which path integration yields $Z_2$. Note that sheets $1$ and $4$ are connected at the right interval $B$. \red{(e) Smooth deformation of the surface in (d), which shows topological equivalence to a torus. The blue spheres stand for the blue sheets in (d) and the colored cylinders stand for the connections between the same area of the two neighboring sheets. The yellow dots are the points at infinity of the sheets.  } }
\end{figure}

\emph{A replica approach}---
Following previous works \cite{Hayden2016tensor,liu2022detecting,cft_09rev},
$\mathcal{E}$ can be computed via a ``replica trick'' method (See SM for a rigorous proof): \begin{equation}\label{eq:E=Z}
    \mathcal{E} = \lim_{n\rightarrow 1/2}\ln Z_n, \quad \mathrm{where}\quad Z_n\equiv \tr\left[(\RR^\dagger \RR)^n\right],
\end{equation}
and $\displaystyle\lim_{n\to 1/2}$ means analytic continuation from integer values of $n$ to $\frac{1}{2}$. $Z_n$ can be expressed as contracting $2n$ copies of $\rho$ as tensors (see \cref{fig:Rn}(a) and (b)).
Using imaginary time path integral, any matrix element of $\rho$ for the subsystem $A\cup B$ in the ground state equals the partition function on a 2d plane $\mathbb{C}$ with open cuts at the two intervals, as shown in \cref{fig:Rn}(c). The boundary conditions at the cuts correspond to the four states $|a\rangle,|a'\rangle,|b\rangle,|b'\rangle$ specified by the given matrix element. Connecting the matrix elements according to \cref{fig:Rn}(b), $Z_n$ is then the partition function on a Riemann surface $\mathcal{R}_n$ depicted in \cref{fig:Rn}(d). \red{The most important observation of this paper is that, as the complex plane $\mathbb{C}$ is topologically equivalent to a sphere when compactified, $\mathcal{R}_n$ is topologically equivalent to a \emph{torus} for any value of $n$. We take the case $n=2$ as an example to shown this equivalence in \cref{fig:Rn}(e). This is the key property that make the CCNR negativity easier to calculate than the PPT negativity in CFT.}

When compressed to a single plane, $Z_n$ can be further viewed as the correlation function of some \textit{twist fields} $\TT'_{2n}(z)$ in $2n$ copies of the original theory \cite{cardy2008_twist}
\begin{equation}\label{eq:4pt}
    Z_n = \alr{\TT'_{2n}(u_a)\TT'_{2n}(v_a)\TTT'_{2n}(u_b)\TTT'_{2n}(v_b) },
\end{equation}
where the fields locate at the four end points of $A$ and $B$.
In a nutshell, each sheet of $\mathcal{R}_n$ corresponds to a flavor (labeled by $1,2,\cdots,2n$) in the compressed plane, and $\TT'_{2n}$ and $\TTT'_{2n}$ permute the flavors by $(1\leftrightarrow 2, 3\leftrightarrow 4,\cdots,2n-1\leftrightarrow 2n)$ and $(2\leftrightarrow 3, \cdots,2n\leftrightarrow 1)$ respectively. Note that these twist fields differ from $\TT_{2n}$ for calculating EE and PPT negativity in the literature \cite{cft_09rev,cft_neg_12}, where the permutation is cyclic: $(1\rightarrow 2, 2\rightarrow 3,\cdots,2n\rightarrow 1)$. This replica approach also works for general systems beyond 2d CFT.

\emph{Two adjacent intervals.}---  In 2d CFT, the Riemann surface $\mathcal{R}_n$ can be conformally transformed to more tractable geometries, with well-known transformation properties of primary fields such as $\TT'_{2n}$. As a warm-up, consider the case $v_a=u_b$ so that $A$ and $B$ are adjacent. Then $Z_n$ corresponds to a three-point function $ Z_n = \alr{\TT'_{2n}(u_a)\TT^{\otimes 2}_n(u_b)\TTT'_{2n}(v_b) }$,
where $\TT^{\otimes 2}_n$, the composition of $\TT'_{2n}$ and $\TTT'_{2n}$, permutes the flavors by $(1\rightarrow 3, 3\rightarrow5, \cdots,2n-1\rightarrow 1)$ and $(2\rightarrow 2n, 4\rightarrow 2, \cdots,2n\rightarrow 2n-2)$. This justifies the notation $\TT^{\otimes 2}_n$, which means the odd and even groups of flavors factorize, and there is a cyclic permutation in each group. In CFT, three-point functions take a universal form \footnote{We refer to \cite{CFT_book} for CFT basics used in this Letter.} that only depends on the geometry, the central charge $c$, and the conformal dimensions of the three operators that we compute as follows. To obtain the conformal dimension $h_{\TT'_{2n}}=\bar{h}_{\TT'_{2n}}$ for $\TT'_{2n}$ (the dimension for $\TTT'_{2n}$ would be the same), consider the two-point function $\alr{\TT'_{2n}(u) \TT'_{2n}(v)}\sim |u-v|^{-4h_{\TT'_{2n}}}$. The corresponding Riemann surface is $n$ independent copies of the $n=1$ case, where two sheets are connected by a cut linking $u$ to $v$, so that $\TT'_2 = \TT_2$. Therefore we have 
\begin{equation}\label{eq:h'}
    h_{\TT'_{2n}} = n h_{\TT_2} = n\frac{c}{24}\lr{2-\frac{1}{2}} = \frac{n}{16}c,
\end{equation}
where we use the well-known value of $h_{\TT_n}$ \cite{cft_09rev}. Similarly, we have $h_{\TT^{\otimes 2}_n} = 2h_{\TT_n} = \frac{c}{12}\lr{n-1/n}$.

As a result, we find \begin{align}\label{eq:Zn_3pt}
    Z_n &\propto (\ell_a\ell_b)^{-2h_{\TT^{\otimes 2}_n}} (\ell_a+\ell_b)^{2h_{\TT^{\otimes 2}_n} - 4 h_{\TT'_{2n}}} \nonumber\\
    &= (\ell_a\ell_b)^{-\frac{c}{6}\lr{n-\frac{1}{n}}} (\ell_a+\ell_b)^{-\frac{c}{12}\lr{n+\frac{2}{n}}}.
\end{align}
In the limit $n\rightarrow 1/2$, we get for two adjacent intervals \begin{equation}\label{eq:adja}
    \mathcal{E} = \frac{c}{8}\mlr{2\ln(\ell_a\ell_b)-3\ln(\ell_a+\ell_b)}+\mathrm{const}.
\end{equation}
Using standard CFT techniques, this result can be easily generalized to finite size or finite temperature \cite{cft_09rev}. For example, if the system is of length $L$ with periodic boundary condition, $\mathcal{E}$ at zero temperature is still given by \cref{eq:adja}, but with each length $\ell$ replaced by $\frac{L}{\pi}\sin\frac{\pi \ell}{L}$.

\emph{Two disjoint intervals.}--- 
If $A$ and $B$ are disjoint, we should use the four-point function \cref{eq:4pt}, which can be rewritten as 
\begin{equation}\label{eq:Z=F}
    Z_n = \lr{\frac{|u_a-u_b||v_a-v_b|}{\ell_a\ell_b|u_a-v_b||u_b-v_a|}}^{\frac{nc}{4}}\mathcal{F}_{2n}(x)
\end{equation}
using global conformal transformations and the conformal dimension in \cref{eq:h'}. Here the four-point ratio $x$ is given by \cref{eq:x},
and the function \begin{align}\label{eq:Fn}
    \mathcal{F}_{2n}(x) =|x(1-x)|^{\frac{nc}{4}}\alr{\TT'_{2n}(0)\TT'_{2n}(x)\TTT'_{2n}(1)\TTT'_{2n}(\infty) },
\end{align} 
is proportional to $Z_n$ defined at $(u_a,v_a,u_b,v_b)=(0,x,1,\infty)$. From now on, we focus on this particular geometry, with the operator at $\infty$ normalized by $\TTT'_{2n}(\infty)=\lim_{w\to \infty} |w|^{\frac{nc}{4}}\TTT'_{2n}(w)$. The subscript $2n$ makes $\mathcal{F}_2(x)$ agree with previous notations \cite{2int_09,2int_11,cft_neg_12}, where the two-sheet Riemann surface for calculating EE or PPT negativity is exactly the same as CCNR negativity here.
$\mathcal{F}_{2n}(x)$ is not universal and depends on the full operator content of the theory since the topology of the Riemann surface $\mathcal{R}_n$ is no longer a plane (strictly speaking, a sphere). However, the topology is just a little more complicated than a plane: it is a \emph{torus for all $n$} (see \cref{fig:Rn}(d)). This special property about $\mathcal{E}$, which does not hold for EE and PPT negativity, enables us to derive universal relations between entanglement and finite temperature physics described by a torus.

\begin{figure}[t]
\includegraphics[width=0.45\textwidth]{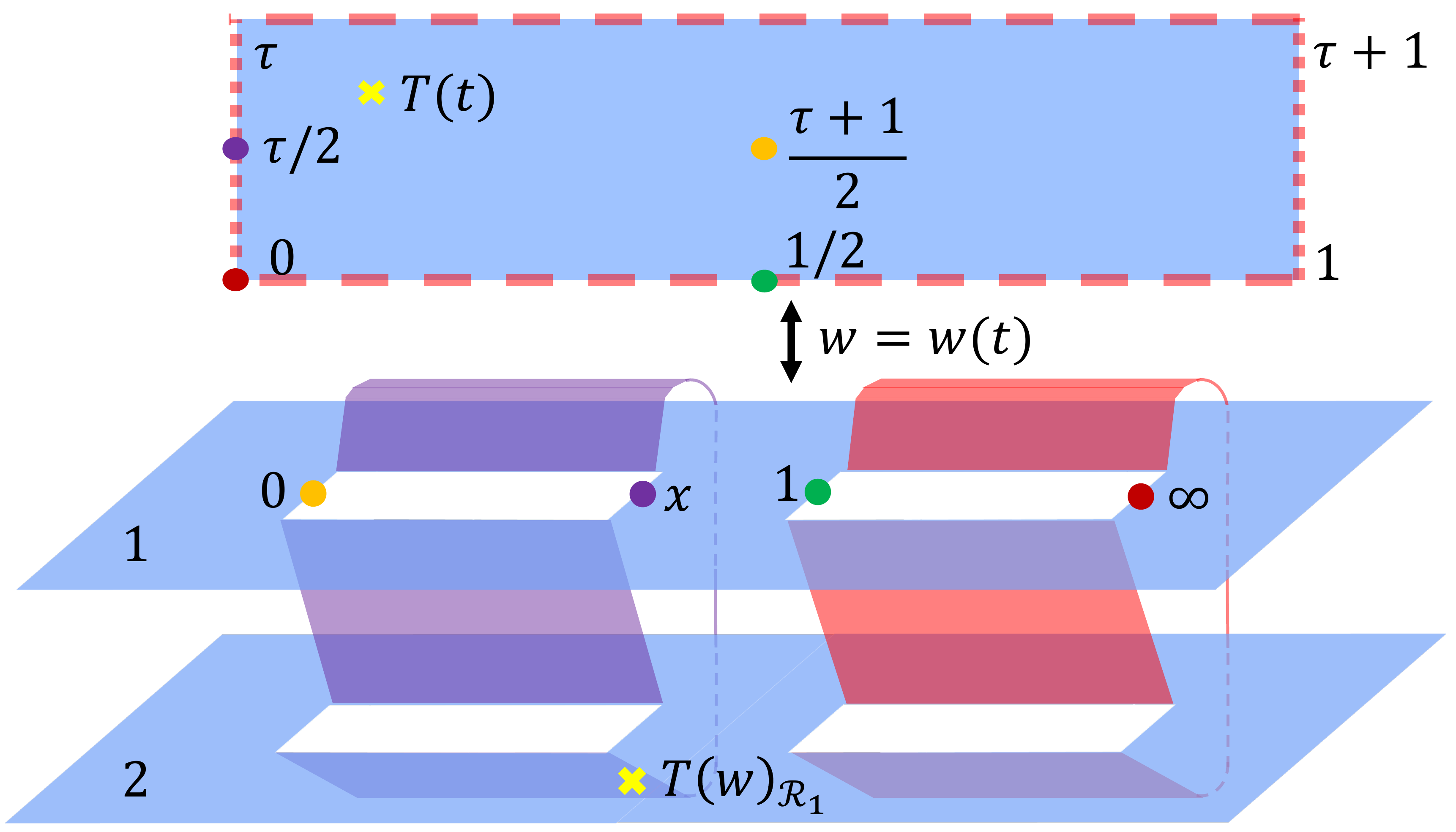}
\caption{\label{fig:map}  Schematic depiction of the one-to-one mapping in \cref{eq:wt} between the torus $T_\tau$ above, represented by a rectangle with opposite sides identified, and $\mathcal{R}_1$ below. Points with the same color are mapped to each other.}
\end{figure}

We show the universal relation by first considering the simplest case $n=1$, where we introduce our main technique depicted in \cref{fig:map}. Namely, there is a one-to-one mapping between the Riemann surface $\mathcal{R}_1$ and a torus $T_\tau$, first introduced in \cite{DIXON1987}. We parametrize $\mathcal{R}_1$ by $w\in \mathbb{C}$ with one value of $w$ corresponding to two points in $\mathcal{R}_1$ (except for the four end points of $A$ and $B$). On the other hand, the torus $T_\tau$ is defined by the coordinate $t\in \mathbb{C}$ with periodic identifications $t\cong t+p+q\tau$, where $p$ and $q$ are integers. Here $\tau$ is the modular parameter determined by \cref{eq:xt}. Using this parametrization, the map is written as
\begin{equation}\label{eq:wt}
    w(t) = \frac{\wp(t)-e_3}{e_1-e_3}, 
\end{equation} 
where $\wp(t)$ is the Weierstrass elliptic function on a lattice generated by $1$ and $\tau$ \footnote{We refer to \cite{NIST:DLMF,DIXON1987} for details on the special functions used in this Letter.}, and $e_1,e_2,e_3$ equal to $\wp(1/2),\wp(\tau/2),\wp((1+\tau)/2)$ respectively with constraint  \begin{equation}\label{eq:e=0}
    e_1+e_2+e_3=0.
\end{equation} 
$w(t)$ maps $T_\tau$ one-to-two to the complex plane, except for the four points $t=(1+\tau)/2, 1/2, 0,\tau/2$ that map to the four end points $w=0,1,\infty,x$ respectively, due to \cref{eq:xt}.

To obtain $\mathcal{F}_2(x)$, we insert a stress tensor $T(w)$ in \cref{eq:Fn} and calculate the five-point function first \footnote{This strategy follows from the CFT calculation for EE of a single interval \cite{cft_04,cft_09rev}}. This is equivalent to a single-point function of the stress tensor $T(w)_{\mathcal{R}_1}$ on $\mathcal{R}_1$ with an extra prefactor $2$, since it has two sheets. According to the map in \cref{eq:wt}, this is then related to the single-point function of $T(t)$ on $T_\tau$ from the transformation rule \begin{equation}\label{eq:T=T}
    T(w)_{\mathcal{R}_1} = \lr{\ddd{w}{t}}^{-2} \lr{T(t)-\frac{c}{12}\glr{w,t} },
\end{equation}
where $\glr{w,t} =w'''/w'-\frac{3}{2}(w''/w')^2 = \wp'''/\wp'-\frac{3}{2}(\wp''/\wp')^2$ is the Schwarzian derivative. To simplify, observe that $\wp'' = \wp'\ddd{\wp'}{\wp} = 6 \lr{\wp^2+ \epsilon }$,
where $3\epsilon\equiv e_1e_2+e_2e_3+e_3e_1$, and we have used the identity \begin{equation}\label{eq:p'=p}
    \wp'^2 = 4\lr{\wp-e_1}\lr{\wp-e_2}\lr{\wp-e_3},
\end{equation} 
together with \cref{eq:e=0}.
Then $\wp'''=12\wp\wp'$ follows, and we get \begin{equation}\label{eq:wt=p}
    \frac{1}{12}\glr{w,t} = \wp(t) - \frac{9\lr{\wp^2+ \epsilon }^2}{8\lr{\wp-e_1}\lr{\wp-e_2}\lr{\wp-e_3}},
\end{equation}
for the second term in \cref{eq:T=T}.
For the first term, \red{ we derive its expectation in SM:} \begin{equation}\label{eq:T=dZ}
    \alr{T(t)}_{T_\tau} = 2\pi\ii\partial_\tau \ln Z(\tau).
\end{equation}

Taking the expectation value of \cref{eq:T=T}, we obtain \begin{align}\label{eq:5pt}
    &\alr{T(w)\TT'_2(0)\TT'_2(x)\TTT'_2(1)\TTT'_2(\infty)} \nonumber\\ &= 2\alr{T(w)_{\mathcal{R}_1}}_{\mathcal{R}_1} = \lr{\frac{e_1-e_3}{\wp'(t)}}^2 \lr{2\alr{T(t)}_{T_\tau} -\frac{c}{6}\glr{w,t}} \nonumber\\ &= \frac{1}{w-x} \lr{\frac{\alr{T(t)}_{T_\tau}}{2(e_1-e_3)x(x-1)} - \frac{c}{24}\frac{2x-1}{x(x-1)} }+\cdots.
\end{align}
In the third line we have used \cref{eq:p'=p,eq:wt=p} and extracted the pole of order $1$ at $w=x$. According to the conformal Ward identity, the residue of the five-point function \cref{eq:5pt} should equal to $\partial_x \alr{\TT'_2(0)\TT'_2(x)\TTT'_2(1)\TTT'_2(\infty)}$. Using the identity \begin{equation}\label{eq:dxdt}
    2(e_1-e_3)x(x-1) = -2\pi^2 x\theta_4(\tau)^4 = 2\pi\ii \ddd{x}{\tau},
\end{equation}
and \cref{eq:Fn}, we then integrate over $x$ to get \begin{equation}\label{eq:F2}
    \mathcal{F}_{2}(x) = Z(\tau) \ilr{x(1-x)}^{\frac{c}{6}}.
\end{equation}
This establishes a universal relation between the R\' enyi-2 EE (or equivalently, purity) $S_2=-\ln Z_1$ of two disjoint intervals and the torus partition function.
As an example, Ref.~\cite{2int_info_09} reports $\mathcal{F}_2(x)$ for the free compactified boson (CB) model with a critical exponent $\eta$. This is easily reproduced using \cref{eq:F2} and the partition function \cite{CFT_book} \begin{equation}\label{eq:Z_CB}
    Z_{\mathrm{CB}}(\tau) = \sqrt{\frac{\eta}{-\ii\tau}}\frac{\theta_3(\eta\tau)\theta_3(-\eta/\tau)}{\mlr{\theta_2(\tau)\theta_3(\tau)\theta_4(\tau)}^{2/3}}.
\end{equation}

Thanks to the torus topology, we generalize the calculation for all $n\ge 1$ in SM, where the odd (even) sheets in $\mathcal{R}_n$ are compressed to the up (down) sheet in \cref{fig:map}, so that we can still use \cref{eq:wt}. We obtain our main result
\begin{equation}\label{eq:Zn}
    Z_n = \frac{ Z\lr{n\tau}}{\lr{\ell_a\ell_b|u_a-u_b||v_a-v_b||u_a-v_b||u_b-v_a|}^{nc/12}},
\end{equation}
and \cref{eq:main}, with simplified formulas for the two limits $x\rightarrow 0,1$ reported in SM.
As \cref{eq:Zn} provides an \emph{infinite} number of exact constraints on the state $\rho$, it is an interesting question what useful information beyond the CCNR negativity and purity that one can extract from the $Z_n$s. In SM we give a first attempt, according to the natural connection between the matrix $\RR$ and the correlation function $\tr\left[(\OO_A\otimes \OO_B)\rho\right]=\bra{\OO_A^*}\RR\ket{\OO_B}$, where $\ket{\OO_A}$ and $\ket{\OO_B}$ are the vectorizations of operators $\OO_A$ and $\OO_B$, respectively. Thus, according to the Cauchy-Schwarz inequality, we find that $Z_{n\rightarrow \infty}$ bounds the correlation function of low-rank Hermitian operators $\OO_A,\OO_B$ as
\begin{equation}
    \tr\left[(\OO_A\otimes \OO_B)\rho\right]\lesssim \lim_{n\rightarrow\infty}\ (Z_n)^{\frac{1}{2n}} \propto (\ell_a\ell_b)^{-c/8}.
\end{equation}

\begin{figure}
    \centering
    \includegraphics[width=0.46\textwidth]{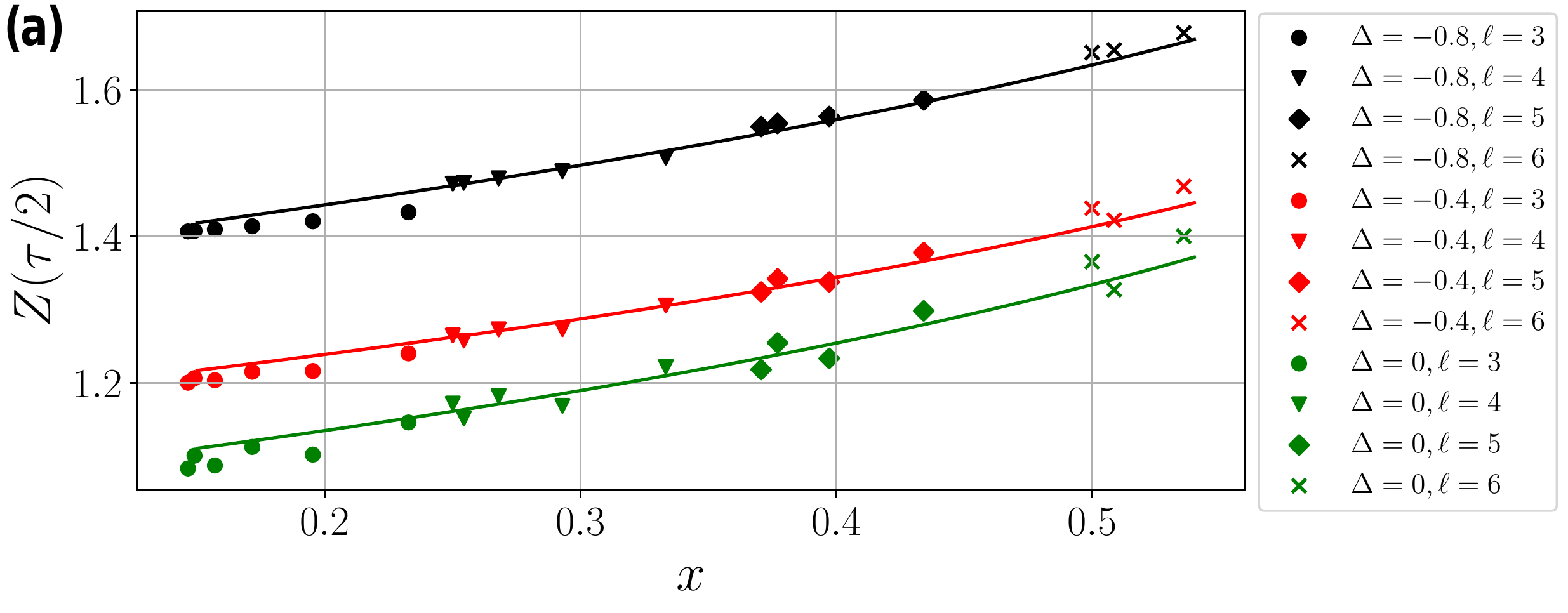}
    \includegraphics[width=0.225\textwidth]{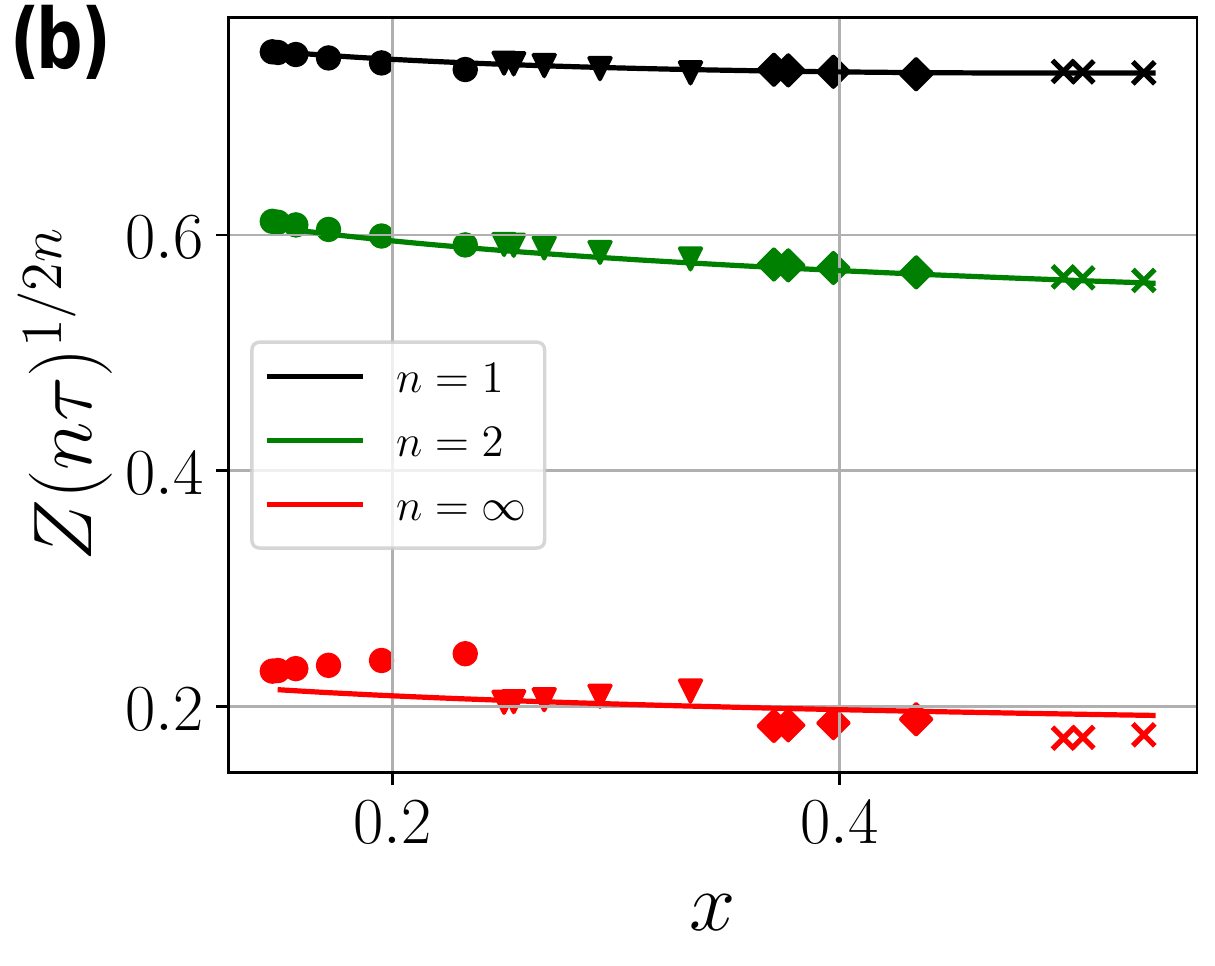}
    \includegraphics[width=0.225\textwidth]{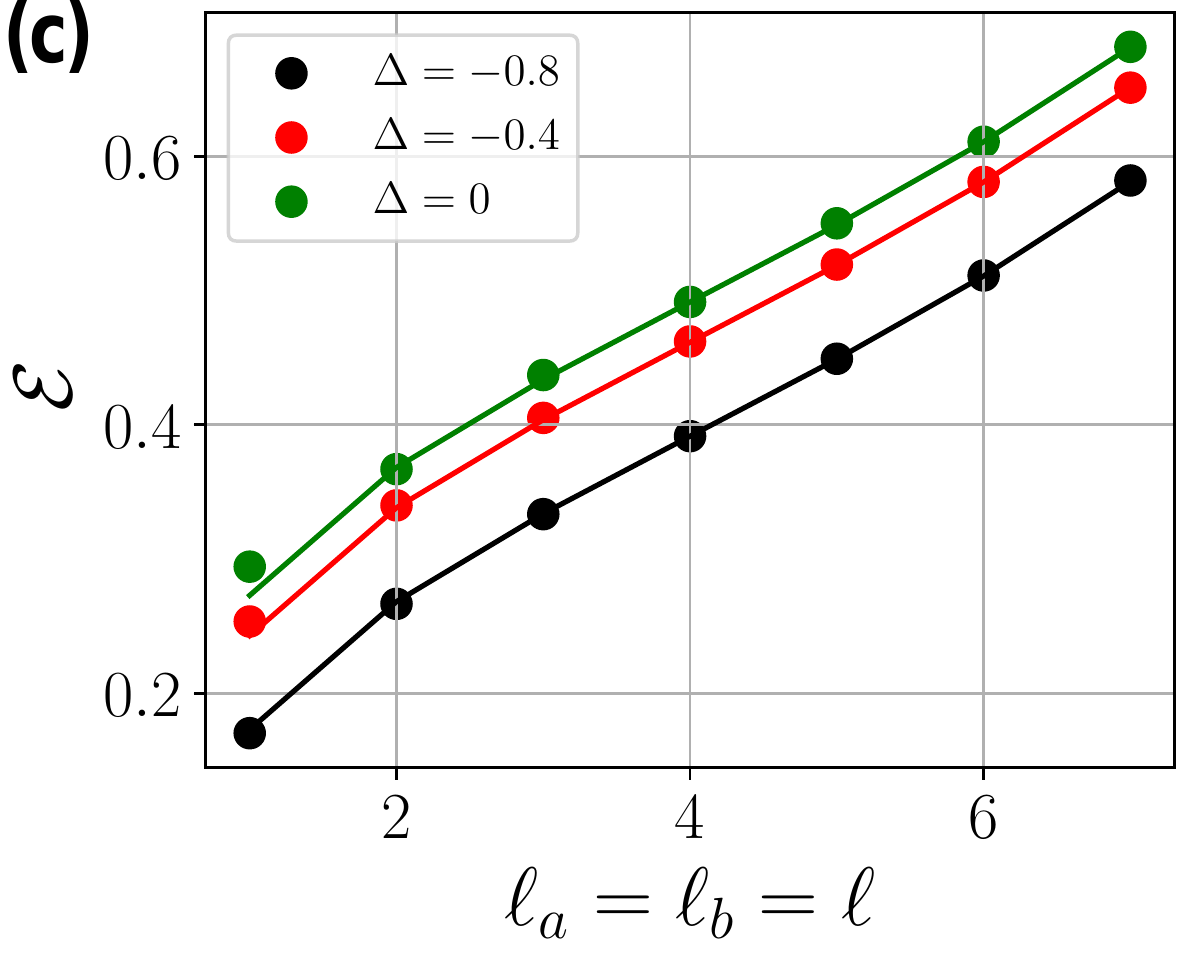}
    \caption{\label{fig:num} Predicted values using \cref{eq:adja,eq:Z_CB} (lines) versus real values (markers) for two intervals with length $\ell_a=\ell_b=\ell$ in the ground state of the $24$-qubit XXZ chain. (a) The real values come from multiplying the numerical CCNR negativity $\ee^{\mathcal{E}}$ of two disjoint intervals with the denominator in \cref{eq:main}. The geometry is determined by the four-point ratio $x$ and $\ell$ indicated by the marker symbol. The three colors stand for different values of $\Delta$. (b) The real values come from \cref{eq:Zn} using the numerical $R$ matrix for different $n$ and a fixed $\Delta=-0.8$. The geometries are the same as (a). (c) CCNR negativity of two adjacent intervals, compared with the prediction \cref{eq:adja}. }
\end{figure}

\emph{Numerics.}--- We use the spin-$1/2$ XXZ chain with periodic boundaries 
\begin{equation}
H = \sum_{j=1}^L X_jX_{j+1} + Y_jY_{j+1} + \Delta Z_jZ_{j+1}
\end{equation}
to test our findings, where the ground state is described by the CFT of a free compactified boson (equivalently, the Luttinger liquid) with $c=1$ and critical exponent $\eta = 1-\frac{1}{\pi}\arccos\Delta$ \cite{2int_info_09}. We numerically calculate the ground state for $L=24$ sites by exact diagonalization and extract the $R$ matrix and CCNR negativity for different geometries and values of $\Delta$. As shown in \cref{fig:num}(a) and (c), the data agrees well with our predictions \cref{eq:adja,eq:main} using \cref{eq:Z_CB} for the partition function. In \cref{fig:num}(b), the general formula \cref{eq:Zn} is also verified for $\Delta=-0.8$.

\emph{Discussion.}---
In conclusion, we discover that the entanglement of two disjoint intervals in ($1+1$)-d CFTs, as quantified by CCNR negativity, is universally related to the thermal partition function. Furthermore, similar relations hold for the R\' enyi counterparts $Z_n$ that provide extra information about the state $\rho$, such as the purity and a bound on correlation function. Our work thus adds to a series of rigorous findings on many-body problems \cite{tripart_1D,neg_info_dual,MT_L,entan_spread_area}, where it is crucial to choose the suitable entanglement measures that echo with the particular many-body structure.

We expect our results can be generalized in many directions, such as going beyond 1d ground states to excited states \cite{cft_excitation_11,cft_excited_12} and finite temperature \cite{cft_T_14} at higher dimensions \cite{cft_2d_06,cft_corner_15}. The quantity $Z_n$ naturally appears in the replica trick for the reflected entropy \cite{reflected_19,reflected_21,reflected_22}, which is nicely dual to the entanglement wedge cross section \cite{wedge_cross} in AdS/CFT. Thus it is worth exploring the meaning of \cref{eq:Zn} in holographic settings, \red{see \cite{ccn_holo22} for a recent discussion}. Since our main results can be alternatively viewed as solving four-point functions of twist fields, it is interesting to ask whether a similar structure holds for disorder operators \cite{diso_71,diso_17,diso_DQC,diso_21}, the generalization of twist field operators in the symmetry perspective. 

As one more generalization, one can ask about entanglement and correlation for $N>2$ intervals. Our result for the two-interval purity already yields the
R\' enyi-2 $N$-partite information \cite{n_info}, for $N=3$ intervals where at least two are adjacent, and $N=4$ adjacent intervals. For example, the R\' enyi-2 tripartite information for intervals $A,B,C$ is 
\begin{align}
    I_2(A:B:C) &= S_2(A) + S_2(B) + S_2(C) - S_2(AB)\nonumber\\ &\quad - S_2(AC) - S_2(BC) + S_2(ABC),
\end{align}
which only contains purities for one or two intervals, if $A$ is adjacent to $B$. On the other hand, for any $N$, one can construct families of Riemann surfaces that are topologically a torus, such as connecting each pair of neighboring sheets by only one interval. However, it is an open question whether our technique \cref{eq:wt} can be generalized to such Riemann surfaces. It is also unclear whether these Riemann surfaces lead to meaningful measures of entanglement and correlation.

\emph{Acknowledgements.}--- We thank Andrew Lucas and Xiaoliang Qi for valuable comments. We thank Thomas Faulkner and Pratik Rath for informing us the connection to the reflected entropy. This work was supported by the National Natural Science Foundation of China Grants No.~12174216.

\onecolumngrid
\newpage
\twocolumngrid

\renewcommand{\theequation}{S\arabic{equation}}
\renewcommand{\thefigure}{S\arabic{figure}}
\setcounter{equation}{0}
\setcounter{figure}{0}

\begin{center}
{\large \textbf{Supplementary Material}}
\end{center}

\section{justification of the replica approach}
To see \cref{eq:E=Z}, we write $\ee^{\mathcal{E}} = f(1/2)$, where the function $f(\zeta) \equiv \sum_j \lambda_j^{\zeta}$ with $\lambda_j>0$ is the $j$th largest eigenvalue of $\RR^\dagger \RR$. We also have $\lambda_j\le 1$ because their sum $f(1)=\tr{\RR^\dagger \RR}=\tr{\rho^2}\le 1$. As a result, as long as $f(1/2)$ is finite, $f(\zeta)$ will be uniformly convergent, and thus analytic, in the region $\mathrm{Re}\zeta\ge 1/2$. Therefore, $f(1/2)$ can be analytically continued from the values $f(n)=Z_n$ when $\zeta=n$ is a positive integer.

\red{
\section{expectation of the stress tensor on a torus}
Consider a torus $T_\tau$ with modular parameter $\tau$, as shown in the upper part of \cref{fig:map}. We calculate the expectation $\alr{T(t)}_{T_\tau}$ of the stress tensor.
First, we use $t=\frac{\ii}{2\pi}\ln z$ to map the $z$-plane to the $t$-cylinder, and get \begin{equation}
    \alr{T(t)}_{T_\tau} = Z(\tau)^{-1}\mtr{T(t) q^{L_0-c/24}\bar{q}^{\bar{L}_0-c/24}} \label{eq:T=tr},
\end{equation}
where $q=\ee^{2\pi\ii\tau}$, its complex conjugate $\bar{q}=\ee^{-2\pi\ii\bar{\tau}}$, $L_m,\bar{L}_m$ are the Virasoro generators, and the partition function on $T_\tau$ is \begin{equation}\label{eq:Zt}
    Z(\tau) = \mtr{q^{L_0-c/24}\bar{q}^{\bar{L}_0-c/24}}.
\end{equation}
Then, the transformation $T(t) = -(2\pi)^2 \lr{\sum_m z^{-m} L_m - \frac{c}{24}}$
from \cref{eq:T=T} contributes to \cref{eq:T=tr} only by the $L_0$ and $c$ terms, which yields the result \cref{eq:T=dZ}.
}

\section{derivation for $Z_{n>1}$}
\begin{figure}[t]
\includegraphics[width=0.45\textwidth]{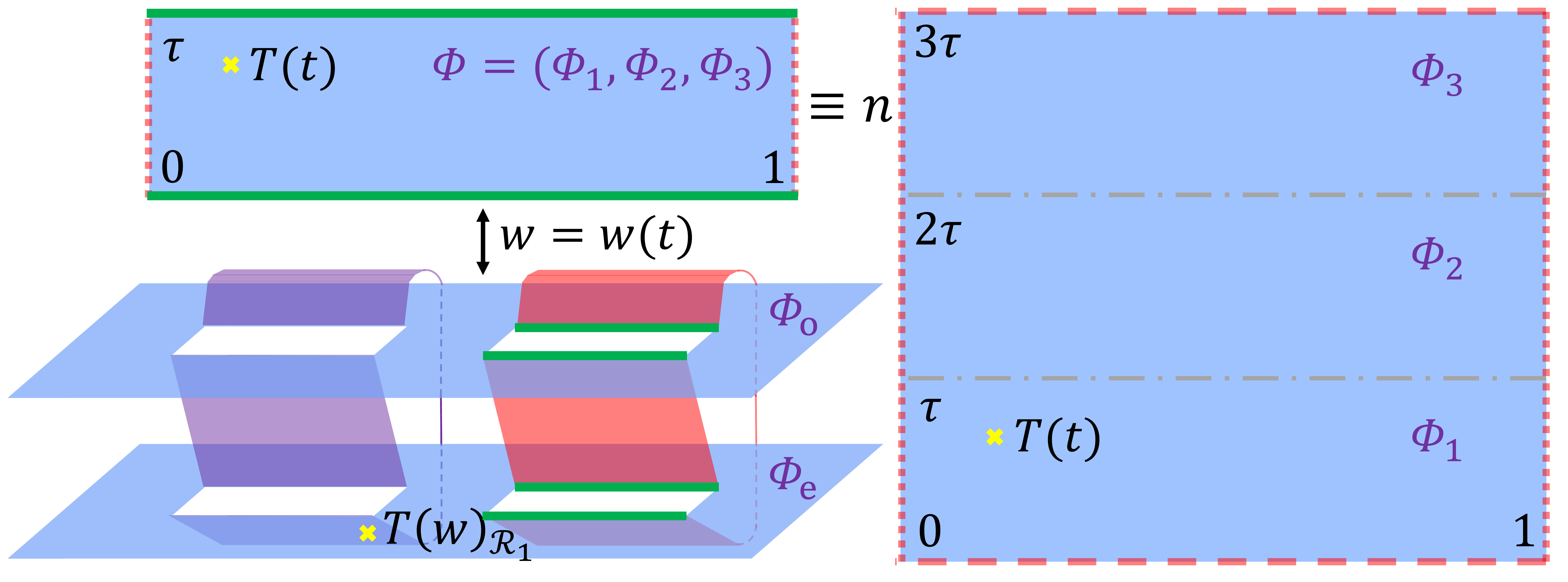}
\caption{\label{fig:map1}  For $n>1$, e.g. $n=3$, \cref{eq:wt} maps the five-point function to the torus $T^n_{\tau,\mathrm{tw}}$ with $n$ copies of the quantum field, and twisted boundary condition indicated by the green solid line. This is further mapped to $T_{n\tau}$ with periodic boundary conditions, with a factor $n$ in front. }
\end{figure}

Observe that $Z_n$ is equivalent to the partition function on $\mathcal{R}_1$ with $n$ flavors on each of the two sheets: The $n$ flavors of the first (second) sheet correspond to the fields on the odd (even) sheets of $\mathcal{R}_n$. Denoting the field on the $j$th sheet of $\mathcal{R}_n$ by $\phi_j$, the field on the first sheet of $\mathcal{R}_1$ is then $\Phi_\oo=(\Phi_{\oo,1},\cdots, \Phi_{\oo,n})\equiv (\phi_1,\phi_3,\cdots,\phi_{2n-1})$, while that on the second is $\Phi_\ee=(\Phi_{\ee,1},\cdots, \Phi_{\ee,n})\equiv (\phi_2,\phi_4,\cdots,\phi_{2n})$. One also need to specify the continuity conditions at the two intervals that connect the two sheets of $\mathcal{R}_1$: The condition at $A$ is normal: $\Phi_\oo=\Phi_\ee$, while that at $B$ involves a cyclic permutation: $\Phi_{\oo,j}=\Phi_{\ee,j-1}, (j=1,\cdots,n)$ where subscript $0\equiv n$ (see \cref{fig:map1}). As a consequence, \cref{eq:wt} maps this theory on $\mathcal{R}_1$ to an $n$-component field $\Phi=(\Phi_1,\cdots,\Phi_n)$ living on the torus with modular parameter $\tau$, and a twisted boundary condition at the circle $\mathrm{Im}t=0$: $\Phi_j(t+\ii 0)=\Phi_{j-1}(t-\ii 0)$. We denote this torus by $T^n_{\tau,\mathrm{tw}}$ to emphasize it contains $n$ copies of the original CFT, and the boundary condition is twisted. 
Following the previous derivation in \cref{eq:5pt}, we get \begin{align}\label{eq:5ptn}
    &\alr{T(w)\TT'_{2n}(0)\TT'_{2n}(x)\TTT'_{2n}(1)\TTT'_{2n}(\infty)} \nonumber\\ &= \frac{1}{w-x} \lr{\frac{\alr{T(t)}_{T^n_{\tau,\mathrm{tw}}}}{2(e_1-e_3)x(x-1)} - \frac{nc}{24}\frac{2x-1}{x(x-1)} }+\cdots,
\end{align}
where $nc$ is the central charge for $n$ copies of the original CFT.
By writing down the path integral explicitly, the theory on $T^n_{\tau,\mathrm{tw}}$ can be unfolded as a single copy of the original CFT, on the $n$-times elongated torus $T_{n\tau}$ with no twist, as depicted in \cref{fig:map1}. Using \cref{eq:T=dZ}, the stress tensor is then \begin{equation}
    \alr{T(t)}_{T^n_{\tau,\mathrm{tw}}} = n\alr{T(t)}_{T_{n\tau}} = 2\pi\ii\partial_\tau \ln Z(n\tau),
\end{equation}
Following the derivation around \cref{eq:dxdt}, we get
\begin{equation}\label{eq:F=Z}
    \mathcal{F}_{2n}(x) = Z(n\tau)\ilr{x(x-1)}^{\frac{nc}{6}}.
\end{equation}
Combining \cref{eq:x,eq:E=Z,eq:Z=F,eq:F=Z}, we arrive at \cref{eq:Zn}.

\section{close and far intervals limits}
In the limit where $A$ and $B$ are far or close to each other with respect to their lengths, \cref{eq:Zn} (and therefore \cref{eq:main}) reduces to a universal function independent of the function $Z(\tau)$. In the far limit, we have $x\approx 16\ee^{\pi\ii\tau}\to 0$ from \cref{eq:xt}. Using $L_0,\bar{L}_0\ge 0$, \cref{eq:Zt} implies \begin{equation}\label{eq:Zt'}
    Z(\tau')\propto \mathrm{exp}(-\pi\ii c\tau'/6) \propto x^{-nc/6},
\end{equation}
where $\tau'=n\tau\rightarrow \ii\infty$, and \cref{eq:Zn} becomes \begin{equation} \label{eq:n_inf}
    Z_n\propto \lr{\ell_a\ell_b}^{-nc/4},
\end{equation}
Thus $\mathcal{E}$ approaches a constant at large separation $s=u_b-v_a$. 

In the close limit $s\to 0$, we verify that \cref{eq:Zn} reduces to \cref{eq:Zn_3pt} for adjacent intervals. Here $1-x\approx (\ell_a+\ell_b)s/(\ell_a\ell_b)\to 0$, and $\tau\to 0$ with $1-x\propto \mathrm{exp}(-\pi\ii/\tau)$ from \cref{eq:xt}. Then from modular invariance $Z(-1/\tau')=Z(\tau')$ and asymptotics in \cref{eq:Zt'}, we have $Z(n\tau)\propto (1-x)^{-c/(6n)}$. \cref{eq:Zn_3pt} is then reproduced from \cref{eq:Zn}, with an extra factor $s^{-c(n+2/n)/12}$ where $s$ should be set to (some multiples of) the underlying lattice spacing.

\section{$Z_\infty$ bounds correlation function}
\begin{figure}
    \centering
    \includegraphics[width=0.45\textwidth]{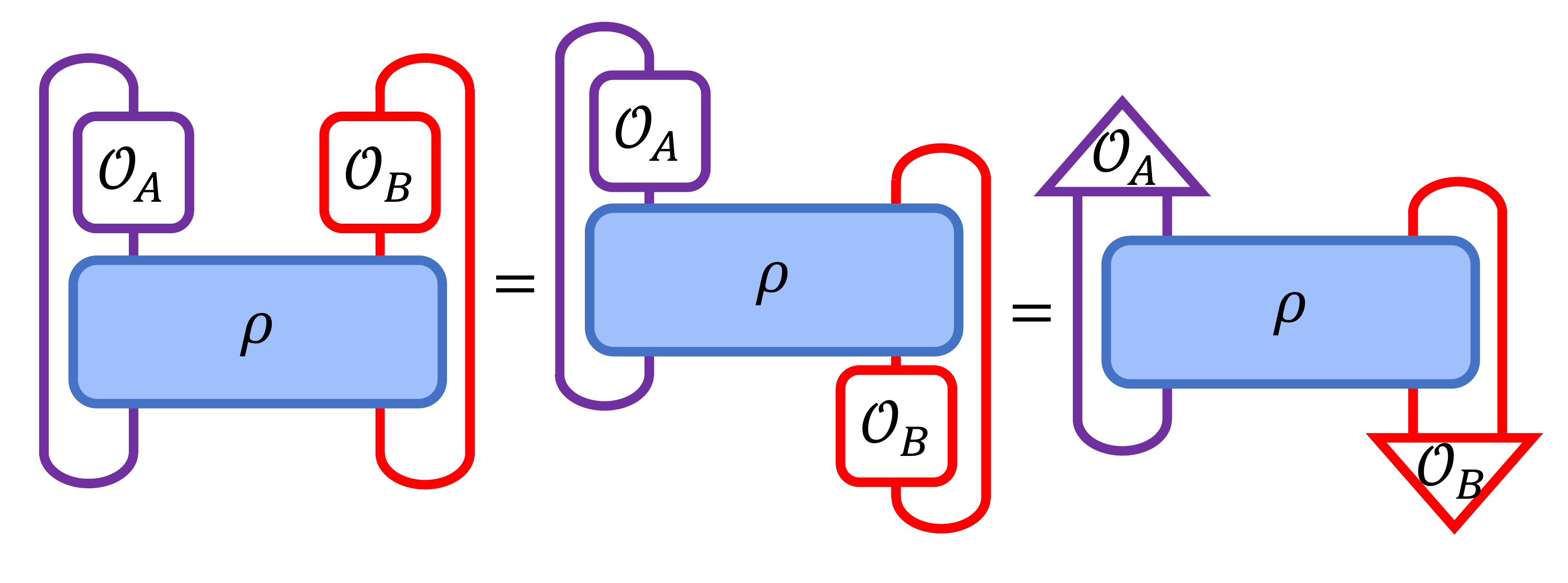}
    \caption{Relation between the $R$ matrix used in CCNR and correlation function, where operators $\OO_A,\OO_B$ can be viewed as two vectors indicated by the triangles. }
    \label{fig:cor}
\end{figure}

As shown in \cref{fig:cor}, we write
\begin{align}\label{eq:cor<R}
    &\quad \tr\left[(\OO_A\otimes \OO_B)\rho\right] = \alr{\OO_A^*|R|\OO_B}\nonumber\\ 
    &\le \sqrt{\alr{\OO_A^* | \OO_A^*} \alr{\OO_B|\OO_B}}\ \norm{R}_\infty \nonumber\\ &= \sqrt{ \tr{\OO_A^2}\tr{\OO_B^2} } \lim_{n\rightarrow\infty}\ (Z_n)^{\frac{1}{2n}} \nonumber\\
    &\propto \sqrt{ \tr{\OO_A^2}\tr{\OO_B^2} }\ (\ell_a\ell_b)^{-c/8}.
\end{align}
Here in the first line, we have used \eqref{eq:RR} and defined $|\OO_B\rangle = \sum_{bb'}\lr{\OO_B}_{bb'}|b\rangle|b'\rangle$ and similarly for $\langle \OO_A^*|$. The second line follows from Cauchy-Schwarz inequality, where $\norm{\cdot}_\infty$ is the operator norm. The third line follows by definition of the states and \cref{eq:E=Z}, and the last line uses \cref{eq:n_inf} which holds whenever $n\tau\rightarrow \ii \infty$. In order for the bound \cref{eq:cor<R} to be meaningful, we require $\OO_A$ (and $\OO_B$) to be low-rank in the sense that $\tr{\OO_A^2}/ \norm{\OO_A}^2_\infty$ grows as a sufficiently slow power law with $\ell_a$, so that \cref{eq:cor<R} beats the trivial bound $\tr\left[(\OO_A\otimes \OO_B)\rho\right] \le \norm{\OO_A}_\infty \norm{\OO_B}_\infty$. For such operators, \cref{eq:cor<R} predicts that the correlation does not depend on the separation $s$, and decays as a power law with the lengths of the two intervals, with a universal decay exponent $c/8$.

\bibliography{CCNR_CFT}

\providecommand{\noopsort}[1]{}\providecommand{\singleletter}[1]{#1}%
\begin{thebibliography}{57}%
\makeatletter
\providecommand \@ifxundefined [1]{%
 \@ifx{#1\undefined}
}%
\providecommand \@ifnum [1]{%
 \ifnum #1\expandafter \@firstoftwo
 \else \expandafter \@secondoftwo
 \fi
}%
\providecommand \@ifx [1]{%
 \ifx #1\expandafter \@firstoftwo
 \else \expandafter \@secondoftwo
 \fi
}%
\providecommand \natexlab [1]{#1}%
\providecommand \enquote  [1]{``#1''}%
\providecommand \bibnamefont  [1]{#1}%
\providecommand \bibfnamefont [1]{#1}%
\providecommand \citenamefont [1]{#1}%
\providecommand \href@noop [0]{\@secondoftwo}%
\providecommand \href [0]{\begingroup \@sanitize@url \@href}%
\providecommand \@href[1]{\@@startlink{#1}\@@href}%
\providecommand \@@href[1]{\endgroup#1\@@endlink}%
\providecommand \@sanitize@url [0]{\catcode `\\12\catcode `\$12\catcode
  `\&12\catcode `\#12\catcode `\^12\catcode `\_12\catcode `\%12\relax}%
\providecommand \@@startlink[1]{}%
\providecommand \@@endlink[0]{}%
\providecommand \url  [0]{\begingroup\@sanitize@url \@url }%
\providecommand \@url [1]{\endgroup\@href {#1}{\urlprefix }}%
\providecommand \urlprefix  [0]{URL }%
\providecommand \Eprint [0]{\href }%
\providecommand \doibase [0]{http://dx.doi.org/}%
\providecommand \selectlanguage [0]{\@gobble}%
\providecommand \bibinfo  [0]{\@secondoftwo}%
\providecommand \bibfield  [0]{\@secondoftwo}%
\providecommand \translation [1]{[#1]}%
\providecommand \BibitemOpen [0]{}%
\providecommand \bibitemStop [0]{}%
\providecommand \bibitemNoStop [0]{.\EOS\space}%
\providecommand \EOS [0]{\spacefactor3000\relax}%
\providecommand \BibitemShut  [1]{\csname bibitem#1\endcsname}%
\let\auto@bib@innerbib\@empty
\bibitem [{\citenamefont {Amico}\ \emph {et~al.}(2008)\citenamefont {Amico},
  \citenamefont {Fazio}, \citenamefont {Osterloh},\ and\ \citenamefont
  {Vedral}}]{entan_many}%
  \BibitemOpen
  \bibfield  {author} {\bibinfo {author} {\bibfnamefont {Luigi}\ \bibnamefont
  {Amico}}, \bibinfo {author} {\bibfnamefont {Rosario}\ \bibnamefont {Fazio}},
  \bibinfo {author} {\bibfnamefont {Andreas}\ \bibnamefont {Osterloh}}, \ and\
  \bibinfo {author} {\bibfnamefont {Vlatko}\ \bibnamefont {Vedral}},\
  }\bibfield  {title} {\enquote {\bibinfo {title} {Entanglement in many-body
  systems},}\ }\href {\doibase 10.1103/RevModPhys.80.517} {\bibfield  {journal}
  {\bibinfo  {journal} {Rev. Mod. Phys.}\ }\textbf {\bibinfo {volume} {80}},\
  \bibinfo {pages} {517--576} (\bibinfo {year} {2008})}\BibitemShut {NoStop}%
\bibitem [{\citenamefont {Calabrese}\ and\ \citenamefont
  {Cardy}(2004)}]{cft_04}%
  \BibitemOpen
  \bibfield  {author} {\bibinfo {author} {\bibfnamefont {Pasquale}\
  \bibnamefont {Calabrese}}\ and\ \bibinfo {author} {\bibfnamefont {John}\
  \bibnamefont {Cardy}},\ }\bibfield  {title} {\enquote {\bibinfo {title}
  {Entanglement entropy and quantum field theory},}\ }\href {\doibase
  10.1088/1742-5468/2004/06/p06002} {\bibfield  {journal} {\bibinfo  {journal}
  {Journal of Statistical Mechanics: Theory and Experiment}\ }\textbf {\bibinfo
  {volume} {2004}},\ \bibinfo {pages} {P06002} (\bibinfo {year}
  {2004})}\BibitemShut {NoStop}%
\bibitem [{\citenamefont {Calabrese}\ and\ \citenamefont
  {Cardy}(2009)}]{cft_09rev}%
  \BibitemOpen
  \bibfield  {author} {\bibinfo {author} {\bibfnamefont {Pasquale}\
  \bibnamefont {Calabrese}}\ and\ \bibinfo {author} {\bibfnamefont {John}\
  \bibnamefont {Cardy}},\ }\bibfield  {title} {\enquote {\bibinfo {title}
  {Entanglement entropy and conformal field theory},}\ }\href {\doibase
  10.1088/1751-8113/42/50/504005} {\bibfield  {journal} {\bibinfo  {journal}
  {Journal of Physics A: Mathematical and Theoretical}\ }\textbf {\bibinfo
  {volume} {42}},\ \bibinfo {pages} {504005} (\bibinfo {year}
  {2009})}\BibitemShut {NoStop}%
\bibitem [{\citenamefont {Fradkin}\ and\ \citenamefont
  {Moore}(2006)}]{cft_2d_06}%
  \BibitemOpen
  \bibfield  {author} {\bibinfo {author} {\bibfnamefont {Eduardo}\ \bibnamefont
  {Fradkin}}\ and\ \bibinfo {author} {\bibfnamefont {Joel~E.}\ \bibnamefont
  {Moore}},\ }\bibfield  {title} {\enquote {\bibinfo {title} {Entanglement
  entropy of 2d conformal quantum critical points: Hearing the shape of a
  quantum drum},}\ }\href {\doibase 10.1103/PhysRevLett.97.050404} {\bibfield
  {journal} {\bibinfo  {journal} {Phys. Rev. Lett.}\ }\textbf {\bibinfo
  {volume} {97}},\ \bibinfo {pages} {050404} (\bibinfo {year}
  {2006})}\BibitemShut {NoStop}%
\bibitem [{\citenamefont {Calabrese}\ \emph {et~al.}(2010)\citenamefont
  {Calabrese}, \citenamefont {Campostrini}, \citenamefont {Essler},\ and\
  \citenamefont {Nienhuis}}]{cft_sublead_10}%
  \BibitemOpen
  \bibfield  {author} {\bibinfo {author} {\bibfnamefont {Pasquale}\
  \bibnamefont {Calabrese}}, \bibinfo {author} {\bibfnamefont {Massimo}\
  \bibnamefont {Campostrini}}, \bibinfo {author} {\bibfnamefont {Fabian}\
  \bibnamefont {Essler}}, \ and\ \bibinfo {author} {\bibfnamefont {Bernard}\
  \bibnamefont {Nienhuis}},\ }\bibfield  {title} {\enquote {\bibinfo {title}
  {Parity effects in the scaling of block entanglement in gapless spin
  chains},}\ }\href {\doibase 10.1103/PhysRevLett.104.095701} {\bibfield
  {journal} {\bibinfo  {journal} {Phys. Rev. Lett.}\ }\textbf {\bibinfo
  {volume} {104}},\ \bibinfo {pages} {095701} (\bibinfo {year}
  {2010})}\BibitemShut {NoStop}%
\bibitem [{\citenamefont {Alcaraz}\ \emph {et~al.}(2011)\citenamefont
  {Alcaraz}, \citenamefont {Berganza},\ and\ \citenamefont
  {Sierra}}]{cft_excitation_11}%
  \BibitemOpen
  \bibfield  {author} {\bibinfo {author} {\bibfnamefont {Francisco~Castilho}\
  \bibnamefont {Alcaraz}}, \bibinfo {author} {\bibfnamefont {Miguel
  Ib\'a\~nez}\ \bibnamefont {Berganza}}, \ and\ \bibinfo {author}
  {\bibfnamefont {Germ\'an}\ \bibnamefont {Sierra}},\ }\bibfield  {title}
  {\enquote {\bibinfo {title} {Entanglement of low-energy excitations in
  conformal field theory},}\ }\href {\doibase 10.1103/PhysRevLett.106.201601}
  {\bibfield  {journal} {\bibinfo  {journal} {Phys. Rev. Lett.}\ }\textbf
  {\bibinfo {volume} {106}},\ \bibinfo {pages} {201601} (\bibinfo {year}
  {2011})}\BibitemShut {NoStop}%
\bibitem [{\citenamefont {Calabrese}\ \emph {et~al.}(2012)\citenamefont
  {Calabrese}, \citenamefont {Cardy},\ and\ \citenamefont
  {Tonni}}]{cft_neg_12}%
  \BibitemOpen
  \bibfield  {author} {\bibinfo {author} {\bibfnamefont {Pasquale}\
  \bibnamefont {Calabrese}}, \bibinfo {author} {\bibfnamefont {John}\
  \bibnamefont {Cardy}}, \ and\ \bibinfo {author} {\bibfnamefont {Erik}\
  \bibnamefont {Tonni}},\ }\bibfield  {title} {\enquote {\bibinfo {title}
  {Entanglement negativity in quantum field theory},}\ }\href {\doibase
  10.1103/PhysRevLett.109.130502} {\bibfield  {journal} {\bibinfo  {journal}
  {Phys. Rev. Lett.}\ }\textbf {\bibinfo {volume} {109}},\ \bibinfo {pages}
  {130502} (\bibinfo {year} {2012})}\BibitemShut {NoStop}%
\bibitem [{\citenamefont {Berganza}\ \emph {et~al.}(2012)\citenamefont
  {Berganza}, \citenamefont {Alcaraz},\ and\ \citenamefont
  {Sierra}}]{cft_excited_12}%
  \BibitemOpen
  \bibfield  {author} {\bibinfo {author} {\bibfnamefont
  {Miguel~Ib{\'{a}}{\~{n}}ez}\ \bibnamefont {Berganza}}, \bibinfo {author}
  {\bibfnamefont {Francisco~Castilho}\ \bibnamefont {Alcaraz}}, \ and\ \bibinfo
  {author} {\bibfnamefont {Germ{\'{a}}n}\ \bibnamefont {Sierra}},\ }\bibfield
  {title} {\enquote {\bibinfo {title} {Entanglement of excited states in
  critical spin chains},}\ }\href {\doibase 10.1088/1742-5468/2012/01/p01016}
  {\bibfield  {journal} {\bibinfo  {journal} {Journal of Statistical Mechanics:
  Theory and Experiment}\ }\textbf {\bibinfo {volume} {2012}},\ \bibinfo
  {pages} {P01016} (\bibinfo {year} {2012})}\BibitemShut {NoStop}%
\bibitem [{\citenamefont {Cardy}\ and\ \citenamefont
  {Herzog}(2014)}]{cft_T_14}%
  \BibitemOpen
  \bibfield  {author} {\bibinfo {author} {\bibfnamefont {John}\ \bibnamefont
  {Cardy}}\ and\ \bibinfo {author} {\bibfnamefont {Christopher~P.}\
  \bibnamefont {Herzog}},\ }\bibfield  {title} {\enquote {\bibinfo {title}
  {Universal thermal corrections to single interval entanglement entropy for
  two dimensional conformal field theories},}\ }\href {\doibase
  10.1103/PhysRevLett.112.171603} {\bibfield  {journal} {\bibinfo  {journal}
  {Phys. Rev. Lett.}\ }\textbf {\bibinfo {volume} {112}},\ \bibinfo {pages}
  {171603} (\bibinfo {year} {2014})}\BibitemShut {NoStop}%
\bibitem [{\citenamefont {Bueno}\ \emph {et~al.}(2015)\citenamefont {Bueno},
  \citenamefont {Myers},\ and\ \citenamefont {Witczak-Krempa}}]{cft_corner_15}%
  \BibitemOpen
  \bibfield  {author} {\bibinfo {author} {\bibfnamefont {Pablo}\ \bibnamefont
  {Bueno}}, \bibinfo {author} {\bibfnamefont {Robert~C.}\ \bibnamefont
  {Myers}}, \ and\ \bibinfo {author} {\bibfnamefont {William}\ \bibnamefont
  {Witczak-Krempa}},\ }\bibfield  {title} {\enquote {\bibinfo {title}
  {Universality of corner entanglement in conformal field theories},}\ }\href
  {\doibase 10.1103/PhysRevLett.115.021602} {\bibfield  {journal} {\bibinfo
  {journal} {Phys. Rev. Lett.}\ }\textbf {\bibinfo {volume} {115}},\ \bibinfo
  {pages} {021602} (\bibinfo {year} {2015})}\BibitemShut {NoStop}%
\bibitem [{\citenamefont {Goldstein}\ and\ \citenamefont
  {Sela}(2018)}]{cft_sym_18}%
  \BibitemOpen
  \bibfield  {author} {\bibinfo {author} {\bibfnamefont {Moshe}\ \bibnamefont
  {Goldstein}}\ and\ \bibinfo {author} {\bibfnamefont {Eran}\ \bibnamefont
  {Sela}},\ }\bibfield  {title} {\enquote {\bibinfo {title} {Symmetry-resolved
  entanglement in many-body systems},}\ }\href {\doibase
  10.1103/PhysRevLett.120.200602} {\bibfield  {journal} {\bibinfo  {journal}
  {Phys. Rev. Lett.}\ }\textbf {\bibinfo {volume} {120}},\ \bibinfo {pages}
  {200602} (\bibinfo {year} {2018})}\BibitemShut {NoStop}%
\bibitem [{\citenamefont {Caraglio}\ and\ \citenamefont
  {Gliozzi}(2008)}]{2int_08}%
  \BibitemOpen
  \bibfield  {author} {\bibinfo {author} {\bibfnamefont {Michele}\ \bibnamefont
  {Caraglio}}\ and\ \bibinfo {author} {\bibfnamefont {Ferdinando}\ \bibnamefont
  {Gliozzi}},\ }\bibfield  {title} {\enquote {\bibinfo {title} {Entanglement
  entropy and twist fields},}\ }\href {\doibase 10.1088/1126-6708/2008/11/076}
  {\bibfield  {journal} {\bibinfo  {journal} {Journal of High Energy Physics}\
  }\textbf {\bibinfo {volume} {2008}},\ \bibinfo {pages} {076} (\bibinfo {year}
  {2008})}\BibitemShut {NoStop}%
\bibitem [{\citenamefont {Furukawa}\ \emph {et~al.}(2009)\citenamefont
  {Furukawa}, \citenamefont {Pasquier},\ and\ \citenamefont
  {Shiraishi}}]{2int_info_09}%
  \BibitemOpen
  \bibfield  {author} {\bibinfo {author} {\bibfnamefont {Shunsuke}\
  \bibnamefont {Furukawa}}, \bibinfo {author} {\bibfnamefont {Vincent}\
  \bibnamefont {Pasquier}}, \ and\ \bibinfo {author} {\bibfnamefont {Jun'ichi}\
  \bibnamefont {Shiraishi}},\ }\bibfield  {title} {\enquote {\bibinfo {title}
  {Mutual information and boson radius in a $c=1$ critical system in one
  dimension},}\ }\href {\doibase 10.1103/PhysRevLett.102.170602} {\bibfield
  {journal} {\bibinfo  {journal} {Phys. Rev. Lett.}\ }\textbf {\bibinfo
  {volume} {102}},\ \bibinfo {pages} {170602} (\bibinfo {year}
  {2009})}\BibitemShut {NoStop}%
\bibitem [{\citenamefont {Calabrese}\ \emph {et~al.}(2009)\citenamefont
  {Calabrese}, \citenamefont {Cardy},\ and\ \citenamefont {Tonni}}]{2int_09}%
  \BibitemOpen
  \bibfield  {author} {\bibinfo {author} {\bibfnamefont {Pasquale}\
  \bibnamefont {Calabrese}}, \bibinfo {author} {\bibfnamefont {John}\
  \bibnamefont {Cardy}}, \ and\ \bibinfo {author} {\bibfnamefont {Erik}\
  \bibnamefont {Tonni}},\ }\bibfield  {title} {\enquote {\bibinfo {title}
  {Entanglement entropy of two disjoint intervals in conformal field theory},}\
  }\href {\doibase 10.1088/1742-5468/2009/11/p11001} {\bibfield  {journal}
  {\bibinfo  {journal} {Journal of Statistical Mechanics: Theory and
  Experiment}\ }\textbf {\bibinfo {volume} {2009}},\ \bibinfo {pages} {P11001}
  (\bibinfo {year} {2009})}\BibitemShut {NoStop}%
\bibitem [{\citenamefont {Calabrese}\ \emph {et~al.}(2011)\citenamefont
  {Calabrese}, \citenamefont {Cardy},\ and\ \citenamefont {Tonni}}]{2int_11}%
  \BibitemOpen
  \bibfield  {author} {\bibinfo {author} {\bibfnamefont {Pasquale}\
  \bibnamefont {Calabrese}}, \bibinfo {author} {\bibfnamefont {John}\
  \bibnamefont {Cardy}}, \ and\ \bibinfo {author} {\bibfnamefont {Erik}\
  \bibnamefont {Tonni}},\ }\bibfield  {title} {\enquote {\bibinfo {title}
  {Entanglement entropy of two disjoint intervals in conformal field theory:
  {II}},}\ }\href {\doibase 10.1088/1742-5468/2011/01/p01021} {\bibfield
  {journal} {\bibinfo  {journal} {Journal of Statistical Mechanics: Theory and
  Experiment}\ }\textbf {\bibinfo {volume} {2011}},\ \bibinfo {pages} {P01021}
  (\bibinfo {year} {2011})}\BibitemShut {NoStop}%
\bibitem [{\citenamefont {Horodecki}\ \emph {et~al.}(2009)\citenamefont
  {Horodecki}, \citenamefont {Horodecki}, \citenamefont {Horodecki},\ and\
  \citenamefont {Horodecki}}]{entan_rmp}%
  \BibitemOpen
  \bibfield  {author} {\bibinfo {author} {\bibfnamefont {Ryszard}\ \bibnamefont
  {Horodecki}}, \bibinfo {author} {\bibfnamefont {Pawe\l{}}\ \bibnamefont
  {Horodecki}}, \bibinfo {author} {\bibfnamefont {Micha\l{}}\ \bibnamefont
  {Horodecki}}, \ and\ \bibinfo {author} {\bibfnamefont {Karol}\ \bibnamefont
  {Horodecki}},\ }\bibfield  {title} {\enquote {\bibinfo {title} {Quantum
  entanglement},}\ }\href {\doibase 10.1103/RevModPhys.81.865} {\bibfield
  {journal} {\bibinfo  {journal} {Rev. Mod. Phys.}\ }\textbf {\bibinfo {volume}
  {81}},\ \bibinfo {pages} {865--942} (\bibinfo {year} {2009})}\BibitemShut
  {NoStop}%
\bibitem [{\citenamefont {Alba}\ \emph {et~al.}(2010)\citenamefont {Alba},
  \citenamefont {Tagliacozzo},\ and\ \citenamefont
  {Calabrese}}]{2int_Ising_10}%
  \BibitemOpen
  \bibfield  {author} {\bibinfo {author} {\bibfnamefont {Vincenzo}\
  \bibnamefont {Alba}}, \bibinfo {author} {\bibfnamefont {Luca}\ \bibnamefont
  {Tagliacozzo}}, \ and\ \bibinfo {author} {\bibfnamefont {Pasquale}\
  \bibnamefont {Calabrese}},\ }\bibfield  {title} {\enquote {\bibinfo {title}
  {Entanglement entropy of two disjoint blocks in critical ising models},}\
  }\href {\doibase 10.1103/PhysRevB.81.060411} {\bibfield  {journal} {\bibinfo
  {journal} {Phys. Rev. B}\ }\textbf {\bibinfo {volume} {81}},\ \bibinfo
  {pages} {060411} (\bibinfo {year} {2010})}\BibitemShut {NoStop}%
\bibitem [{\citenamefont {Fagotti}\ and\ \citenamefont
  {Calabrese}(2010)}]{2int_XY_10}%
  \BibitemOpen
  \bibfield  {author} {\bibinfo {author} {\bibfnamefont {Maurizio}\
  \bibnamefont {Fagotti}}\ and\ \bibinfo {author} {\bibfnamefont {Pasquale}\
  \bibnamefont {Calabrese}},\ }\bibfield  {title} {\enquote {\bibinfo {title}
  {Entanglement entropy of two disjoint blocks in xy chains},}\ }\href
  {\doibase 10.1088/1742-5468/2010/04/P04016} {\bibfield  {journal} {\bibinfo
  {journal} {Journal of Statistical Mechanics: Theory and Experiment}\ }\textbf
  {\bibinfo {volume} {2010}},\ \bibinfo {pages} {P04016} (\bibinfo {year}
  {2010})}\BibitemShut {NoStop}%
\bibitem [{\citenamefont {Alba}\ \emph {et~al.}(2011)\citenamefont {Alba},
  \citenamefont {Tagliacozzo},\ and\ \citenamefont
  {Calabrese}}]{2int_boson_11}%
  \BibitemOpen
  \bibfield  {author} {\bibinfo {author} {\bibfnamefont {Vincenzo}\
  \bibnamefont {Alba}}, \bibinfo {author} {\bibfnamefont {Luca}\ \bibnamefont
  {Tagliacozzo}}, \ and\ \bibinfo {author} {\bibfnamefont {Pasquale}\
  \bibnamefont {Calabrese}},\ }\bibfield  {title} {\enquote {\bibinfo {title}
  {Entanglement entropy of two disjoint intervals in c = 1 theories},}\ }\href
  {\doibase 10.1088/1742-5468/2011/06/P06012} {\bibfield  {journal} {\bibinfo
  {journal} {Journal of Statistical Mechanics: Theory and Experiment}\ }\textbf
  {\bibinfo {volume} {2011}},\ \bibinfo {pages} {P06012} (\bibinfo {year}
  {2011})}\BibitemShut {NoStop}%
\bibitem [{\citenamefont {Rajabpour}\ and\ \citenamefont
  {Gliozzi}(2012)}]{2int_fusion_12}%
  \BibitemOpen
  \bibfield  {author} {\bibinfo {author} {\bibfnamefont {M~A}\ \bibnamefont
  {Rajabpour}}\ and\ \bibinfo {author} {\bibfnamefont {F}~\bibnamefont
  {Gliozzi}},\ }\bibfield  {title} {\enquote {\bibinfo {title} {Entanglement
  entropy of two disjoint intervals from fusion algebra of twist fields},}\
  }\href {\doibase 10.1088/1742-5468/2012/02/P02016} {\bibfield  {journal}
  {\bibinfo  {journal} {Journal of Statistical Mechanics: Theory and
  Experiment}\ }\textbf {\bibinfo {volume} {2012}},\ \bibinfo {pages} {P02016}
  (\bibinfo {year} {2012})}\BibitemShut {NoStop}%
\bibitem [{\citenamefont {Coser}\ \emph {et~al.}(2014)\citenamefont {Coser},
  \citenamefont {Tagliacozzo},\ and\ \citenamefont {Tonni}}]{nint_14}%
  \BibitemOpen
  \bibfield  {author} {\bibinfo {author} {\bibfnamefont {Andrea}\ \bibnamefont
  {Coser}}, \bibinfo {author} {\bibfnamefont {Luca}\ \bibnamefont
  {Tagliacozzo}}, \ and\ \bibinfo {author} {\bibfnamefont {Erik}\ \bibnamefont
  {Tonni}},\ }\bibfield  {title} {\enquote {\bibinfo {title} {On rényi
  entropies of disjoint intervals in conformal field theory},}\ }\href
  {\doibase 10.1088/1742-5468/2014/01/P01008} {\bibfield  {journal} {\bibinfo
  {journal} {Journal of Statistical Mechanics: Theory and Experiment}\ }\textbf
  {\bibinfo {volume} {2014}},\ \bibinfo {pages} {P01008} (\bibinfo {year}
  {2014})}\BibitemShut {NoStop}%
\bibitem [{\citenamefont {Nobili}\ \emph {et~al.}(2015)\citenamefont {Nobili},
  \citenamefont {Coser},\ and\ \citenamefont {Tonni}}]{2int_num_15}%
  \BibitemOpen
  \bibfield  {author} {\bibinfo {author} {\bibfnamefont {Cristiano~De}\
  \bibnamefont {Nobili}}, \bibinfo {author} {\bibfnamefont {Andrea}\
  \bibnamefont {Coser}}, \ and\ \bibinfo {author} {\bibfnamefont {Erik}\
  \bibnamefont {Tonni}},\ }\bibfield  {title} {\enquote {\bibinfo {title}
  {Entanglement entropy and negativity of disjoint intervals in cft: some
  numerical extrapolations},}\ }\href {\doibase
  10.1088/1742-5468/2015/06/P06021} {\bibfield  {journal} {\bibinfo  {journal}
  {Journal of Statistical Mechanics: Theory and Experiment}\ }\textbf {\bibinfo
  {volume} {2015}},\ \bibinfo {pages} {P06021} (\bibinfo {year}
  {2015})}\BibitemShut {NoStop}%
\bibitem [{\citenamefont {Coser}\ \emph
  {et~al.}(2016{\natexlab{a}})\citenamefont {Coser}, \citenamefont {Tonni},\
  and\ \citenamefont {Calabrese}}]{2int_neg_16}%
  \BibitemOpen
  \bibfield  {author} {\bibinfo {author} {\bibfnamefont {Andrea}\ \bibnamefont
  {Coser}}, \bibinfo {author} {\bibfnamefont {Erik}\ \bibnamefont {Tonni}}, \
  and\ \bibinfo {author} {\bibfnamefont {Pasquale}\ \bibnamefont {Calabrese}},\
  }\bibfield  {title} {\enquote {\bibinfo {title} {Towards the entanglement
  negativity of two disjoint intervals for a one dimensional free fermion},}\
  }\href {\doibase 10.1088/1742-5468/2016/03/033116} {\bibfield  {journal}
  {\bibinfo  {journal} {Journal of Statistical Mechanics: Theory and
  Experiment}\ }\textbf {\bibinfo {volume} {2016}},\ \bibinfo {pages} {033116}
  (\bibinfo {year} {2016}{\natexlab{a}})}\BibitemShut {NoStop}%
\bibitem [{\citenamefont {Coser}\ \emph
  {et~al.}(2016{\natexlab{b}})\citenamefont {Coser}, \citenamefont {Tonni},\
  and\ \citenamefont {Calabrese}}]{2int_spin_16}%
  \BibitemOpen
  \bibfield  {author} {\bibinfo {author} {\bibfnamefont {Andrea}\ \bibnamefont
  {Coser}}, \bibinfo {author} {\bibfnamefont {Erik}\ \bibnamefont {Tonni}}, \
  and\ \bibinfo {author} {\bibfnamefont {Pasquale}\ \bibnamefont {Calabrese}},\
  }\bibfield  {title} {\enquote {\bibinfo {title} {Spin structures and
  entanglement of two disjoint intervals in conformal field theories},}\ }\href
  {\doibase 10.1088/1742-5468/2016/05/053109} {\bibfield  {journal} {\bibinfo
  {journal} {Journal of Statistical Mechanics: Theory and Experiment}\ }\textbf
  {\bibinfo {volume} {2016}},\ \bibinfo {pages} {053109} (\bibinfo {year}
  {2016}{\natexlab{b}})}\BibitemShut {NoStop}%
\bibitem [{\citenamefont {Ruggiero}\ \emph {et~al.}(2018)\citenamefont
  {Ruggiero}, \citenamefont {Tonni},\ and\ \citenamefont
  {Calabrese}}]{2int_recursion_18}%
  \BibitemOpen
  \bibfield  {author} {\bibinfo {author} {\bibfnamefont {Paola}\ \bibnamefont
  {Ruggiero}}, \bibinfo {author} {\bibfnamefont {Erik}\ \bibnamefont {Tonni}},
  \ and\ \bibinfo {author} {\bibfnamefont {Pasquale}\ \bibnamefont
  {Calabrese}},\ }\bibfield  {title} {\enquote {\bibinfo {title} {Entanglement
  entropy of two disjoint intervals and the recursion formula for conformal
  blocks},}\ }\href {\doibase 10.1088/1742-5468/aae5a8} {\bibfield  {journal}
  {\bibinfo  {journal} {Journal of Statistical Mechanics: Theory and
  Experiment}\ }\textbf {\bibinfo {volume} {2018}},\ \bibinfo {pages} {113101}
  (\bibinfo {year} {2018})}\BibitemShut {NoStop}%
\bibitem [{\citenamefont {Grava}\ \emph {et~al.}(2021)\citenamefont {Grava},
  \citenamefont {Kels},\ and\ \citenamefont {Tonni}}]{2int_CoulombGas_21}%
  \BibitemOpen
  \bibfield  {author} {\bibinfo {author} {\bibfnamefont {Tamara}\ \bibnamefont
  {Grava}}, \bibinfo {author} {\bibfnamefont {Andrew~P.}\ \bibnamefont {Kels}},
  \ and\ \bibinfo {author} {\bibfnamefont {Erik}\ \bibnamefont {Tonni}},\
  }\bibfield  {title} {\enquote {\bibinfo {title} {Entanglement of two disjoint
  intervals in conformal field theory and the 2d coulomb gas on a lattice},}\
  }\href {\doibase 10.1103/PhysRevLett.127.141605} {\bibfield  {journal}
  {\bibinfo  {journal} {Phys. Rev. Lett.}\ }\textbf {\bibinfo {volume} {127}},\
  \bibinfo {pages} {141605} (\bibinfo {year} {2021})}\BibitemShut {NoStop}%
\bibitem [{\citenamefont {Rockwood}(2022)}]{nint_22}%
  \BibitemOpen
  \bibfield  {author} {\bibinfo {author} {\bibfnamefont {Gavin}\ \bibnamefont
  {Rockwood}},\ }\bibfield  {title} {\enquote {\bibinfo {title} {Replicated
  entanglement negativity for disjoint intervals in the ising conformal field
  theory},}\ }\href {\doibase 10.1088/1742-5468/ac873f} {\bibfield  {journal}
  {\bibinfo  {journal} {Journal of Statistical Mechanics: Theory and
  Experiment}\ }\textbf {\bibinfo {volume} {2022}},\ \bibinfo {pages} {083105}
  (\bibinfo {year} {2022})}\BibitemShut {NoStop}%
\bibitem [{\citenamefont {Ares}\ \emph {et~al.}(2022)\citenamefont {Ares},
  \citenamefont {Calabrese}, \citenamefont {Di~Giulio},\ and\ \citenamefont
  {Murciano}}]{2int_sym_22}%
  \BibitemOpen
  \bibfield  {author} {\bibinfo {author} {\bibfnamefont {Filiberto}\
  \bibnamefont {Ares}}, \bibinfo {author} {\bibfnamefont {Pasquale}\
  \bibnamefont {Calabrese}}, \bibinfo {author} {\bibfnamefont {Giuseppe}\
  \bibnamefont {Di~Giulio}}, \ and\ \bibinfo {author} {\bibfnamefont {Sara}\
  \bibnamefont {Murciano}},\ }\bibfield  {title} {\enquote {\bibinfo {title}
  {{Multi-charged moments of two intervals in conformal field theory}},}\
  }\href {\doibase 10.1007/JHEP09(2022)051} {\bibfield  {journal} {\bibinfo
  {journal} {JHEP}\ }\textbf {\bibinfo {volume} {09}},\ \bibinfo {pages} {051}
  (\bibinfo {year} {2022})},\ \Eprint {http://arxiv.org/abs/2206.01534}
  {arXiv:2206.01534 [hep-th]} \BibitemShut {NoStop}%
\bibitem [{\citenamefont {Francesco}\ \emph {et~al.}(2012)\citenamefont
  {Francesco}, \citenamefont {Mathieu},\ and\ \citenamefont
  {Senechal}}]{CFT_book}%
  \BibitemOpen
  \bibfield  {author} {\bibinfo {author} {\bibfnamefont {P.}~\bibnamefont
  {Francesco}}, \bibinfo {author} {\bibfnamefont {P.}~\bibnamefont {Mathieu}},
  \ and\ \bibinfo {author} {\bibfnamefont {D.}~\bibnamefont {Senechal}},\
  }\href {https://books.google.com/books?id=5u7jBwAAQBAJ} {\emph {\bibinfo
  {title} {Conformal Field Theory}}},\ Graduate Texts in Contemporary Physics\
  (\bibinfo  {publisher} {Springer New York},\ \bibinfo {year}
  {2012})\BibitemShut {NoStop}%
\bibitem [{\citenamefont {Chen}\ and\ \citenamefont
  {Wu}(2003)}]{chen2002matrix}%
  \BibitemOpen
  \bibfield  {author} {\bibinfo {author} {\bibfnamefont {Kai}\ \bibnamefont
  {Chen}}\ and\ \bibinfo {author} {\bibfnamefont {Ling-An}\ \bibnamefont
  {Wu}},\ }\bibfield  {title} {\enquote {\bibinfo {title} {A matrix realignment
  method for recognizing entanglement},}\ }\href@noop {} {\bibfield  {journal}
  {\bibinfo  {journal} {Quantum Info. Comput.}\ }\textbf {\bibinfo {volume}
  {3}},\ \bibinfo {pages} {193–202} (\bibinfo {year} {2003})}\BibitemShut
  {NoStop}%
\bibitem [{\citenamefont {Collins}\ and\ \citenamefont
  {Nechita}(2016)}]{collins2016random}%
  \BibitemOpen
  \bibfield  {author} {\bibinfo {author} {\bibfnamefont {Benoit}\ \bibnamefont
  {Collins}}\ and\ \bibinfo {author} {\bibfnamefont {Ion}\ \bibnamefont
  {Nechita}},\ }\bibfield  {title} {\enquote {\bibinfo {title} {Random matrix
  techniques in quantum information theory},}\ }\href
  {https://aip.scitation.org/doi/full/10.1063/1.4936880} {\bibfield  {journal}
  {\bibinfo  {journal} {Journal of Mathematical Physics}\ }\textbf {\bibinfo
  {volume} {57}},\ \bibinfo {pages} {015215} (\bibinfo {year}
  {2016})}\BibitemShut {NoStop}%
\bibitem [{\citenamefont {Peres}(1996)}]{peres1996separability}%
  \BibitemOpen
  \bibfield  {author} {\bibinfo {author} {\bibfnamefont {Asher}\ \bibnamefont
  {Peres}},\ }\bibfield  {title} {\enquote {\bibinfo {title} {Separability
  criterion for density matrices},}\ }\href {\doibase
  10.1103/PhysRevLett.77.1413} {\bibfield  {journal} {\bibinfo  {journal}
  {Phys. Rev. Lett.}\ }\textbf {\bibinfo {volume} {77}},\ \bibinfo {pages}
  {1413--1415} (\bibinfo {year} {1996})}\BibitemShut {NoStop}%
\bibitem [{\citenamefont {Rath}\ \emph {et~al.}(2023)\citenamefont {Rath},
  \citenamefont {Vitale}, \citenamefont {Murciano}, \citenamefont {Votto},
  \citenamefont {Dubail}, \citenamefont {Kueng}, \citenamefont {Branciard},
  \citenamefont {Calabrese},\ and\ \citenamefont {Vermersch}}]{oper_entan_22}%
  \BibitemOpen
  \bibfield  {author} {\bibinfo {author} {\bibfnamefont {Aniket}\ \bibnamefont
  {Rath}}, \bibinfo {author} {\bibfnamefont {Vittorio}\ \bibnamefont {Vitale}},
  \bibinfo {author} {\bibfnamefont {Sara}\ \bibnamefont {Murciano}}, \bibinfo
  {author} {\bibfnamefont {Matteo}\ \bibnamefont {Votto}}, \bibinfo {author}
  {\bibfnamefont {J\'er\^ome}\ \bibnamefont {Dubail}}, \bibinfo {author}
  {\bibfnamefont {Richard}\ \bibnamefont {Kueng}}, \bibinfo {author}
  {\bibfnamefont {Cyril}\ \bibnamefont {Branciard}}, \bibinfo {author}
  {\bibfnamefont {Pasquale}\ \bibnamefont {Calabrese}}, \ and\ \bibinfo
  {author} {\bibfnamefont {Beno\^{\i}t}\ \bibnamefont {Vermersch}},\ }\bibfield
   {title} {\enquote {\bibinfo {title} {Entanglement barrier and its symmetry
  resolution: Theory and experimental observation},}\ }\href {\doibase
  10.1103/PRXQuantum.4.010318} {\bibfield  {journal} {\bibinfo  {journal} {PRX
  Quantum}\ }\textbf {\bibinfo {volume} {4}},\ \bibinfo {pages} {010318}
  (\bibinfo {year} {2023})}\BibitemShut {NoStop}%
\bibitem [{\citenamefont {Zhou}\ and\ \citenamefont
  {Luitz}(2017)}]{oper_entan_17}%
  \BibitemOpen
  \bibfield  {author} {\bibinfo {author} {\bibfnamefont {Tianci}\ \bibnamefont
  {Zhou}}\ and\ \bibinfo {author} {\bibfnamefont {David~J.}\ \bibnamefont
  {Luitz}},\ }\bibfield  {title} {\enquote {\bibinfo {title} {Operator
  entanglement entropy of the time evolution operator in chaotic systems},}\
  }\href {\doibase 10.1103/PhysRevB.95.094206} {\bibfield  {journal} {\bibinfo
  {journal} {Phys. Rev. B}\ }\textbf {\bibinfo {volume} {95}},\ \bibinfo
  {pages} {094206} (\bibinfo {year} {2017})}\BibitemShut {NoStop}%
\bibitem [{Note1()}]{Note1}%
  \BibitemOpen
  \bibinfo {note} {See Supplemental Material for some derivation details and
  additional discussions.}\BibitemShut {Stop}%
\bibitem [{\citenamefont {Hayden}\ \emph {et~al.}(2016)\citenamefont {Hayden},
  \citenamefont {Nezami}, \citenamefont {Qi}, \citenamefont {Thomas},
  \citenamefont {Walter},\ and\ \citenamefont {Yang}}]{Hayden2016tensor}%
  \BibitemOpen
  \bibfield  {author} {\bibinfo {author} {\bibfnamefont {Patrick}\ \bibnamefont
  {Hayden}}, \bibinfo {author} {\bibfnamefont {Sepehr}\ \bibnamefont {Nezami}},
  \bibinfo {author} {\bibfnamefont {Xiao-Liang}\ \bibnamefont {Qi}}, \bibinfo
  {author} {\bibfnamefont {Nathaniel}\ \bibnamefont {Thomas}}, \bibinfo
  {author} {\bibfnamefont {Michael}\ \bibnamefont {Walter}}, \ and\ \bibinfo
  {author} {\bibfnamefont {Zhao}\ \bibnamefont {Yang}},\ }\bibfield  {title}
  {\enquote {\bibinfo {title} {Holographic duality from random tensor
  networks},}\ }\href {\doibase 10.1007/JHEP11(2016)009} {\bibfield  {journal}
  {\bibinfo  {journal} {Journal of High Energy Physics}\ }\textbf {\bibinfo
  {volume} {2016}},\ \bibinfo {pages} {9} (\bibinfo {year} {2016})}\BibitemShut
  {NoStop}%
\bibitem [{\citenamefont {Liu}\ \emph {et~al.}(2022)\citenamefont {Liu},
  \citenamefont {Tang}, \citenamefont {Dai}, \citenamefont {Liu}, \citenamefont
  {Chen},\ and\ \citenamefont {Ma}}]{liu2022detecting}%
  \BibitemOpen
  \bibfield  {author} {\bibinfo {author} {\bibfnamefont {Zhenhuan}\
  \bibnamefont {Liu}}, \bibinfo {author} {\bibfnamefont {Yifan}\ \bibnamefont
  {Tang}}, \bibinfo {author} {\bibfnamefont {Hao}\ \bibnamefont {Dai}},
  \bibinfo {author} {\bibfnamefont {Pengyu}\ \bibnamefont {Liu}}, \bibinfo
  {author} {\bibfnamefont {Shu}\ \bibnamefont {Chen}}, \ and\ \bibinfo {author}
  {\bibfnamefont {Xiongfeng}\ \bibnamefont {Ma}},\ }\bibfield  {title}
  {\enquote {\bibinfo {title} {Detecting entanglement in quantum many-body
  systems via permutation moments},}\ }\href {\doibase
  10.1103/PhysRevLett.129.260501} {\bibfield  {journal} {\bibinfo  {journal}
  {Phys. Rev. Lett.}\ }\textbf {\bibinfo {volume} {129}},\ \bibinfo {pages}
  {260501} (\bibinfo {year} {2022})}\BibitemShut {NoStop}%
\bibitem [{\citenamefont {Cardy}\ \emph {et~al.}(2008)\citenamefont {Cardy},
  \citenamefont {Castro-Alvaredo},\ and\ \citenamefont
  {Doyon}}]{cardy2008_twist}%
  \BibitemOpen
  \bibfield  {author} {\bibinfo {author} {\bibfnamefont {John~L}\ \bibnamefont
  {Cardy}}, \bibinfo {author} {\bibfnamefont {Olalla~A}\ \bibnamefont
  {Castro-Alvaredo}}, \ and\ \bibinfo {author} {\bibfnamefont {Benjamin}\
  \bibnamefont {Doyon}},\ }\bibfield  {title} {\enquote {\bibinfo {title} {Form
  factors of branch-point twist fields in quantum integrable models and
  entanglement entropy},}\ }\href {\doibase
  https://doi.org/10.1007/s10955-007-9422-x} {\bibfield  {journal} {\bibinfo
  {journal} {Journal of Statistical Physics}\ }\textbf {\bibinfo {volume}
  {130}},\ \bibinfo {pages} {129--168} (\bibinfo {year} {2008})}\BibitemShut
  {NoStop}%
\bibitem [{Note2()}]{Note2}%
  \BibitemOpen
  \bibinfo {note} {We refer to \cite {CFT_book} for CFT basics used in this
  Letter.}\BibitemShut {Stop}%
\bibitem [{\citenamefont {Dixon}\ \emph {et~al.}(1987)\citenamefont {Dixon},
  \citenamefont {Friedan}, \citenamefont {Martinec},\ and\ \citenamefont
  {Shenker}}]{DIXON1987}%
  \BibitemOpen
  \bibfield  {author} {\bibinfo {author} {\bibfnamefont {Lance}\ \bibnamefont
  {Dixon}}, \bibinfo {author} {\bibfnamefont {Daniel}\ \bibnamefont {Friedan}},
  \bibinfo {author} {\bibfnamefont {Emil}\ \bibnamefont {Martinec}}, \ and\
  \bibinfo {author} {\bibfnamefont {Stephen}\ \bibnamefont {Shenker}},\
  }\bibfield  {title} {\enquote {\bibinfo {title} {The conformal field theory
  of orbifolds},}\ }\href {\doibase
  https://doi.org/10.1016/0550-3213(87)90676-6} {\bibfield  {journal} {\bibinfo
   {journal} {Nuclear Physics B}\ }\textbf {\bibinfo {volume} {282}},\ \bibinfo
  {pages} {13--73} (\bibinfo {year} {1987})}\BibitemShut {NoStop}%
\bibitem [{Note3()}]{Note3}%
  \BibitemOpen
  \bibinfo {note} {We refer to \cite {NIST:DLMF,DIXON1987} for details on the
  special functions used in this Letter.}\BibitemShut {Stop}%
\bibitem [{Note4()}]{Note4}%
  \BibitemOpen
  \bibinfo {note} {This strategy follows from the CFT calculation for EE of a
  single interval \cite {cft_04,cft_09rev}}\BibitemShut {NoStop}%
\bibitem [{\citenamefont {Zou}\ \emph {et~al.}(2021)\citenamefont {Zou},
  \citenamefont {Siva}, \citenamefont {Soejima}, \citenamefont {Mong},\ and\
  \citenamefont {Zaletel}}]{tripart_1D}%
  \BibitemOpen
  \bibfield  {author} {\bibinfo {author} {\bibfnamefont {Yijian}\ \bibnamefont
  {Zou}}, \bibinfo {author} {\bibfnamefont {Karthik}\ \bibnamefont {Siva}},
  \bibinfo {author} {\bibfnamefont {Tomohiro}\ \bibnamefont {Soejima}},
  \bibinfo {author} {\bibfnamefont {Roger S.~K.}\ \bibnamefont {Mong}}, \ and\
  \bibinfo {author} {\bibfnamefont {Michael~P.}\ \bibnamefont {Zaletel}},\
  }\bibfield  {title} {\enquote {\bibinfo {title} {Universal tripartite
  entanglement in one-dimensional many-body systems},}\ }\href {\doibase
  10.1103/PhysRevLett.126.120501} {\bibfield  {journal} {\bibinfo  {journal}
  {Phys. Rev. Lett.}\ }\textbf {\bibinfo {volume} {126}},\ \bibinfo {pages}
  {120501} (\bibinfo {year} {2021})}\BibitemShut {NoStop}%
\bibitem [{\citenamefont {Bertini}\ \emph {et~al.}(2022)\citenamefont
  {Bertini}, \citenamefont {Klobas},\ and\ \citenamefont {Lu}}]{neg_info_dual}%
  \BibitemOpen
  \bibfield  {author} {\bibinfo {author} {\bibfnamefont {Bruno}\ \bibnamefont
  {Bertini}}, \bibinfo {author} {\bibfnamefont {Katja}\ \bibnamefont {Klobas}},
  \ and\ \bibinfo {author} {\bibfnamefont {Tsung-Cheng}\ \bibnamefont {Lu}},\
  }\bibfield  {title} {\enquote {\bibinfo {title} {Entanglement negativity and
  mutual information after a quantum quench: Exact link from space-time
  duality},}\ }\href {\doibase 10.1103/PhysRevLett.129.140503} {\bibfield
  {journal} {\bibinfo  {journal} {Phys. Rev. Lett.}\ }\textbf {\bibinfo
  {volume} {129}},\ \bibinfo {pages} {140503} (\bibinfo {year}
  {2022})}\BibitemShut {NoStop}%
\bibitem [{\citenamefont {Friedman}\ \emph {et~al.}(2022)\citenamefont
  {Friedman}, \citenamefont {Yin}, \citenamefont {Hong},\ and\ \citenamefont
  {Lucas}}]{MT_L}%
  \BibitemOpen
  \bibfield  {author} {\bibinfo {author} {\bibfnamefont {Aaron~J.}\
  \bibnamefont {Friedman}}, \bibinfo {author} {\bibfnamefont {Chao}\
  \bibnamefont {Yin}}, \bibinfo {author} {\bibfnamefont {Yifan}\ \bibnamefont
  {Hong}}, \ and\ \bibinfo {author} {\bibfnamefont {Andrew}\ \bibnamefont
  {Lucas}},\ }\bibfield  {title} {\enquote {\bibinfo {title} {{Locality and
  error correction in quantum dynamics with measurement}},}\ }\href@noop {} {\
  (\bibinfo {year} {2022})},\ \Eprint {http://arxiv.org/abs/2206.09929}
  {arXiv:2206.09929 [quant-ph]} \BibitemShut {NoStop}%
\bibitem [{\citenamefont {Anshu}\ \emph {et~al.}(2022)\citenamefont {Anshu},
  \citenamefont {Harrow},\ and\ \citenamefont
  {Soleimanifar}}]{entan_spread_area}%
  \BibitemOpen
  \bibfield  {author} {\bibinfo {author} {\bibfnamefont {Anurag}\ \bibnamefont
  {Anshu}}, \bibinfo {author} {\bibfnamefont {Aram~W}\ \bibnamefont {Harrow}},
  \ and\ \bibinfo {author} {\bibfnamefont {Mehdi}\ \bibnamefont
  {Soleimanifar}},\ }\bibfield  {title} {\enquote {\bibinfo {title}
  {Entanglement spread area law in gapped ground states},}\ }\href {\doibase
  https://doi.org/10.1038/s41567-022-01740-7} {\bibfield  {journal} {\bibinfo
  {journal} {Nature Physics}\ ,\ \bibinfo {pages} {1--5}} (\bibinfo {year}
  {2022})}\BibitemShut {NoStop}%
\bibitem [{\citenamefont {Dutta}\ and\ \citenamefont
  {Faulkner}(2021)}]{reflected_19}%
  \BibitemOpen
  \bibfield  {author} {\bibinfo {author} {\bibfnamefont {Souvik}\ \bibnamefont
  {Dutta}}\ and\ \bibinfo {author} {\bibfnamefont {Thomas}\ \bibnamefont
  {Faulkner}},\ }\bibfield  {title} {\enquote {\bibinfo {title} {{A canonical
  purification for the entanglement wedge cross-section}},}\ }\href {\doibase
  10.1007/JHEP03(2021)178} {\bibfield  {journal} {\bibinfo  {journal} {JHEP}\
  }\textbf {\bibinfo {volume} {03}},\ \bibinfo {pages} {178} (\bibinfo {year}
  {2021})},\ \Eprint {http://arxiv.org/abs/1905.00577} {arXiv:1905.00577
  [hep-th]} \BibitemShut {NoStop}%
\bibitem [{\citenamefont {Akers}\ \emph {et~al.}(2022)\citenamefont {Akers},
  \citenamefont {Faulkner}, \citenamefont {Lin},\ and\ \citenamefont
  {Rath}}]{reflected_21}%
  \BibitemOpen
  \bibfield  {author} {\bibinfo {author} {\bibfnamefont {Chris}\ \bibnamefont
  {Akers}}, \bibinfo {author} {\bibfnamefont {Thomas}\ \bibnamefont
  {Faulkner}}, \bibinfo {author} {\bibfnamefont {Simon}\ \bibnamefont {Lin}}, \
  and\ \bibinfo {author} {\bibfnamefont {Pratik}\ \bibnamefont {Rath}},\
  }\bibfield  {title} {\enquote {\bibinfo {title} {{Reflected entropy in random
  tensor networks}},}\ }\href {\doibase 10.1007/JHEP05(2022)162} {\bibfield
  {journal} {\bibinfo  {journal} {JHEP}\ }\textbf {\bibinfo {volume} {05}},\
  \bibinfo {pages} {162} (\bibinfo {year} {2022})},\ \Eprint
  {http://arxiv.org/abs/2112.09122} {arXiv:2112.09122 [hep-th]} \BibitemShut
  {NoStop}%
\bibitem [{\citenamefont {Akers}\ \emph {et~al.}(2023)\citenamefont {Akers},
  \citenamefont {Faulkner}, \citenamefont {Lin},\ and\ \citenamefont
  {Rath}}]{reflected_22}%
  \BibitemOpen
  \bibfield  {author} {\bibinfo {author} {\bibfnamefont {Chris}\ \bibnamefont
  {Akers}}, \bibinfo {author} {\bibfnamefont {Thomas}\ \bibnamefont
  {Faulkner}}, \bibinfo {author} {\bibfnamefont {Simon}\ \bibnamefont {Lin}}, \
  and\ \bibinfo {author} {\bibfnamefont {Pratik}\ \bibnamefont {Rath}},\
  }\bibfield  {title} {\enquote {\bibinfo {title} {{Reflected entropy in random
  tensor networks. Part II. A topological index from canonical
  purification}},}\ }\href {\doibase 10.1007/JHEP01(2023)067} {\bibfield
  {journal} {\bibinfo  {journal} {JHEP}\ }\textbf {\bibinfo {volume} {01}},\
  \bibinfo {pages} {067} (\bibinfo {year} {2023})},\ \Eprint
  {http://arxiv.org/abs/2210.15006} {arXiv:2210.15006 [hep-th]} \BibitemShut
  {NoStop}%
\bibitem [{\citenamefont {Umemoto}\ and\ \citenamefont
  {Takayanagi}(2018)}]{wedge_cross}%
  \BibitemOpen
  \bibfield  {author} {\bibinfo {author} {\bibfnamefont {Koji}\ \bibnamefont
  {Umemoto}}\ and\ \bibinfo {author} {\bibfnamefont {Tadashi}\ \bibnamefont
  {Takayanagi}},\ }\bibfield  {title} {\enquote {\bibinfo {title} {Entanglement
  of purification through holographic duality},}\ }\href {\doibase
  https://doi.org/10.1038/s41567-018-0075-2} {\bibfield  {journal} {\bibinfo
  {journal} {Nature Physics}\ }\textbf {\bibinfo {volume} {14}},\ \bibinfo
  {pages} {573--577} (\bibinfo {year} {2018})}\BibitemShut {NoStop}%
\bibitem [{\citenamefont {Milekhin}\ \emph {et~al.}(2022)\citenamefont
  {Milekhin}, \citenamefont {Rath},\ and\ \citenamefont {Weng}}]{ccn_holo22}%
  \BibitemOpen
  \bibfield  {author} {\bibinfo {author} {\bibfnamefont {Alexey}\ \bibnamefont
  {Milekhin}}, \bibinfo {author} {\bibfnamefont {Pratik}\ \bibnamefont {Rath}},
  \ and\ \bibinfo {author} {\bibfnamefont {Wayne}\ \bibnamefont {Weng}},\
  }\bibfield  {title} {\enquote {\bibinfo {title} {{Computable Cross Norm in
  Tensor Networks and Holography}},}\ }\href@noop {} {\  (\bibinfo {year}
  {2022})},\ \Eprint {http://arxiv.org/abs/2212.11978} {arXiv:2212.11978
  [hep-th]} \BibitemShut {NoStop}%
\bibitem [{\citenamefont {Kadanoff}\ and\ \citenamefont
  {Ceva}(1971)}]{diso_71}%
  \BibitemOpen
  \bibfield  {author} {\bibinfo {author} {\bibfnamefont {Leo~P.}\ \bibnamefont
  {Kadanoff}}\ and\ \bibinfo {author} {\bibfnamefont {Horacio}\ \bibnamefont
  {Ceva}},\ }\bibfield  {title} {\enquote {\bibinfo {title} {Determination of
  an operator algebra for the two-dimensional ising model},}\ }\href {\doibase
  10.1103/PhysRevB.3.3918} {\bibfield  {journal} {\bibinfo  {journal} {Phys.
  Rev. B}\ }\textbf {\bibinfo {volume} {3}},\ \bibinfo {pages} {3918--3939}
  (\bibinfo {year} {1971})}\BibitemShut {NoStop}%
\bibitem [{\citenamefont {Fradkin}(2017)}]{diso_17}%
  \BibitemOpen
  \bibfield  {author} {\bibinfo {author} {\bibfnamefont {Eduardo}\ \bibnamefont
  {Fradkin}},\ }\bibfield  {title} {\enquote {\bibinfo {title} {{Disorder
  Operators and their Descendants}},}\ }\href {\doibase
  10.1007/s10955-017-1737-7} {\bibfield  {journal} {\bibinfo  {journal} {J.
  Statist. Phys.}\ }\textbf {\bibinfo {volume} {167}},\ \bibinfo {pages} {427}
  (\bibinfo {year} {2017})},\ \Eprint {http://arxiv.org/abs/1610.05780}
  {arXiv:1610.05780 [cond-mat.stat-mech]} \BibitemShut {NoStop}%
\bibitem [{\citenamefont {Wang}\ \emph {et~al.}(2022)\citenamefont {Wang},
  \citenamefont {Ma}, \citenamefont {Cheng},\ and\ \citenamefont
  {Meng}}]{diso_DQC}%
  \BibitemOpen
  \bibfield  {author} {\bibinfo {author} {\bibfnamefont {Yan-Cheng}\
  \bibnamefont {Wang}}, \bibinfo {author} {\bibfnamefont {Nvsen}\ \bibnamefont
  {Ma}}, \bibinfo {author} {\bibfnamefont {Meng}\ \bibnamefont {Cheng}}, \ and\
  \bibinfo {author} {\bibfnamefont {Zi~Yang}\ \bibnamefont {Meng}},\ }\bibfield
   {title} {\enquote {\bibinfo {title} {{Scaling of the disorder operator at
  deconfined quantum criticality}},}\ }\href {\doibase
  10.21468/SciPostPhys.13.6.123} {\bibfield  {journal} {\bibinfo  {journal}
  {SciPost Phys.}\ }\textbf {\bibinfo {volume} {13}},\ \bibinfo {pages} {123}
  (\bibinfo {year} {2022})}\BibitemShut {NoStop}%
\bibitem [{\citenamefont {Wang}\ \emph {et~al.}(2021)\citenamefont {Wang},
  \citenamefont {Cheng},\ and\ \citenamefont {Meng}}]{diso_21}%
  \BibitemOpen
  \bibfield  {author} {\bibinfo {author} {\bibfnamefont {Yan-Cheng}\
  \bibnamefont {Wang}}, \bibinfo {author} {\bibfnamefont {Meng}\ \bibnamefont
  {Cheng}}, \ and\ \bibinfo {author} {\bibfnamefont {Zi~Yang}\ \bibnamefont
  {Meng}},\ }\bibfield  {title} {\enquote {\bibinfo {title} {Scaling of the
  disorder operator at $(2+1)d$ u(1) quantum criticality},}\ }\href {\doibase
  10.1103/PhysRevB.104.L081109} {\bibfield  {journal} {\bibinfo  {journal}
  {Phys. Rev. B}\ }\textbf {\bibinfo {volume} {104}},\ \bibinfo {pages}
  {L081109} (\bibinfo {year} {2021})}\BibitemShut {NoStop}%
\bibitem [{\citenamefont {Ag\'on}\ \emph {et~al.}(2022)\citenamefont {Ag\'on},
  \citenamefont {Bueno}, \citenamefont {Lasso~Andino},\ and\ \citenamefont
  {Vilar~L\'opez}}]{n_info}%
  \BibitemOpen
  \bibfield  {author} {\bibinfo {author} {\bibfnamefont {C\'esar~A.}\
  \bibnamefont {Ag\'on}}, \bibinfo {author} {\bibfnamefont {Pablo}\
  \bibnamefont {Bueno}}, \bibinfo {author} {\bibfnamefont {Oscar}\ \bibnamefont
  {Lasso~Andino}}, \ and\ \bibinfo {author} {\bibfnamefont {Alejandro}\
  \bibnamefont {Vilar~L\'opez}},\ }\bibfield  {title} {\enquote {\bibinfo
  {title} {{Aspects of N-partite information in conformal field theories}},}\
  }\href@noop {} {\  (\bibinfo {year} {2022})},\ \Eprint
  {http://arxiv.org/abs/2209.14311} {arXiv:2209.14311 [hep-th]} \BibitemShut
  {NoStop}%
\bibitem [{{\relax DLMF}()}]{NIST:DLMF}%
  \BibitemOpen
  {\relax DLMF},\ \href {http://dlmf.nist.gov/} {\enquote {\bibinfo {title}
  {{\it NIST Digital Library of Mathematical Functions}},}\ }\bibinfo
  {howpublished} {http://dlmf.nist.gov/, Release 1.1.7 of 2022-10-15},\
  \bibinfo {note} {f.~W.~J. Olver, A.~B. {Olde Daalhuis}, D.~W. Lozier, B.~I.
  Schneider, R.~F. Boisvert, C.~W. Clark, B.~R. Miller, B.~V. Saunders, H.~S.
  Cohl, and M.~A. McClain, eds.}\BibitemShut {Stop}%
\end{thebibliography}%

\end{document}